\documentclass[prd,nofootinbib,preprint,superscriptaddress]{revtex4}
\pdfoutput=1

\usepackage{amsmath,amssymb}
\usepackage{epsfig}
\usepackage{graphicx}
\usepackage[usenames,dvipsnames]{color}
\usepackage{subfigure}
\usepackage{slashed}
\usepackage[colorlinks,citecolor=blue]{hyperref}
\DeclareUnicodeCharacter{2212}{-}
\DeclareUnicodeCharacter{202F}{\,}
\usepackage{color}
{}
{}

\begin{document}
	\title{Self-interacting Dark Matter via Right Handed Neutrino Portal}
	
	\author{Debasish Borah}
	\email{dborah@iitg.ac.in}
	\affiliation{Department of Physics, Indian Institute of Technology Guwahati, Assam 781039, India}
	
	\author{Manoranjan Dutta}
	\email{ph18resch11007@iith.ac.in}
	\affiliation{Department of Physics, Indian Institute of Technology Hyderabad, Kandi, Sangareddy 502284, Telangana, India}
	
	\author{Satyabrata Mahapatra}
	\email{ph18resch11001@iith.ac.in}
	\affiliation{Department of Physics, Indian Institute of Technology Hyderabad, Kandi, Sangareddy 502284, Telangana, India}

	\author{Narendra Sahu}
	\email{nsahu@phy.iith.ac.in}
	\affiliation{Department of Physics, Indian Institute of Technology Hyderabad, Kandi, Sangareddy 502284, Telangana, India}
	
	\begin{abstract}
		We propose a self-interacting dark matter (DM) scenario with right handed neutrino (RHN) portal to the standard model (SM). 
		The dark sector consists of a particle DM, assumed to be a Dirac fermion, and a light mediator in terms of a 
		dark Abelian vector boson to give rise to the required velocity dependent self-interactions in agreement with astrophysical observations. 
		Irrespective of thermal or non-thermal production of such a DM, its 
		final relic remains under-abundant due to efficient annihilation rates of DM into light mediators by virtue of large 
		self-interaction coupling. We then show that a feeble portal of DM-SM interaction via RHN offers a possibility to fill the relic 
		deficit of DM via the late decay of RHN. As RHN also arises naturally in seesaw models explaining the origin of light neutrino masses, we outline two UV complete realizations of the minimal setup in terms of scotogenic and gauged $B-L$ frameworks 
		where connection to neutrino mass and other phenomenology like complementary discovery prospects are discussed. 
	\end{abstract}	
	\maketitle
	\section{Introduction}
	\label{intro}
	Presence of dark matter (DM), a non-luminous, non-baryonic form of matter in the present universe in large proportions is supported by 
	irrefutable evidences gathered at several astrophysics and cosmology based experiments over last few decades. While astrophysical experiments 
	reveal the presence of DM locally at different scales like galaxy, clusters \cite{Zwicky:1933gu, Rubin:1970zza, Clowe:2006eq}, cosmological 
	experiments like Planck, WMAP predict around 26.8\% of the present universe to be made up of DM \cite{Zyla:2020zbs, Aghanim:2018eyx}. In terms of density 
	parameter $\Omega_{\rm DM}$ and $h = \text{Hubble Parameter}/(100 \;\text{km}
	~\text{s}^{-1} \text{Mpc}^{-1})$, the present abundance of this
	form of matter, popularly known as dark matter (DM),
	is conventionally reported as \cite{Aghanim:2018eyx}
	\begin{equation}
		\Omega_{\text{DM}} h^2 = 0.120\pm 0.001 
		\label{dm_relic}
	\end{equation}
	at 68\% CL. While astrophysical evidences are based on low redshift measurements, cosmological measurements rely upon the benchmark ${\rm \Lambda CDM}$ model in order to extrapolate their findings at the scale of recombination to the present epoch. In the standard model of cosmology or ${\rm \Lambda CDM}$ model, CDM refers to cold dark matter (indicating collision-less nature of DM) while $\Lambda$ refers to the cosmological constant proposed to be responsible for late acceleration of the universe. Given that DM has a particle origin like visible matter, it is known that none of the Standard Model (SM) particles can satisfy all the criteria of a particle DM candidate. While several beyond standard model (BSM) proposals exist in the literature for DM, none of them have shown up in experiments yet. Perhaps the most popular DM framework is the weakly interacting massive particle (WIMP) paradigm where a DM particle having mass and interactions similar to those around the electroweak scale gives rise to the observed relic after thermal freeze-out, a remarkable coincidence often referred to as the {\it WIMP Miracle} \cite{Kolb:1990vq}.

	Although the standard model of cosmology or ${\rm \Lambda CDM}$ remains in good agreement with large scale structure of the universe, it suffers from several small scale issues like too-big-to-fail, missing satellite and core-cusp problems \cite{Tulin:2017ara, Bullock:2017xww}. One appealing solution to such small scale problems of CDM, proposed by Spergel and Steinhardt \cite{Spergel:1999mh} is known as the self-interacting dark matter (SIDM) paradigm. Earlier studies can be found in \cite{Carlson:1992fn, deLaix:1995vi}. The advantage of SIDM, as opposed to collision-less CDM, is the way it solves the above-mentioned problems at small scales while reproducing the standard CDM halos at large radii to be in agreement with observations. This is possible as the self-interacting scattering rate is proportional to DM density. The required self-interaction is often quantified in terms of cross section to DM mass as $\sigma/m \sim 1 \; {\rm cm}^2/{\rm g} \approx 2 \times 10^{-24} \; {\rm cm}^2/{\rm GeV}$ \cite{Buckley:2009in, Feng:2009hw, Feng:2009mn, Loeb:2010gj, Zavala:2012us, Vogelsberger:2012ku}. Such large self-interacting cross sections of DM can be naturally realized in scenarios where DM has a light mediator. In such a scenario, self-interaction is stronger for smaller DM velocities such that it can have large impact on small scale structure. On the other hand, the velocity dependent self-interaction gets reduced at larger scales (due to large velocities of DM) to be remain consistent with large scale CDM predictions \cite{Buckley:2009in, Feng:2009hw, Feng:2009mn, Loeb:2010gj, Bringmann:2016din, Kaplinghat:2015aga, Aarssen:2012fx, Tulin:2013teo}. While both scalar or vector boson can act like mediators of DM self-interactions, it is more natural to vector mediators as they can be realized within gauge extensions of the SM. For example, Abelian gauge extended model with a vector like fermion DM charged under this symmetry can give rise to simple SIDM scenarios where the same gauge symmetry can also lead to DM stability. Such a dark sector comprising of DM and light mediator can not be completely hidden and there should be some coupling of the mediator with the SM particles which can bring DM and SM sectors to equilibrium in the early universe or lead to the production of DM from the SM bath. The same DM-SM coupling can also be probed at DM direct detection experiments as well \cite{Kaplinghat:2013yxa, DelNobile:2015uua}. Several model building efforts have been made to realize such scenarios. For example, see \cite{Kouvaris:2014uoa, Bernal:2015ova, Kainulainen:2015sva, Hambye:2019tjt, Cirelli:2016rnw, Kahlhoefer:2017umn, Dutta:2021wbn, Borah:2021yek} and references therein.

	We consider a vector like fermion singlet charged under a $U(1)_D$ gauge symmetry as DM. The singlet fermion DM has light mediator interactions in terms of $U(1)_D$ gauge boson $Z'$ having mass which is a few order of magnitudes lighter than DM mass, required for desired SIDM phenomenology. Since the required self-interaction constrains the corresponding gauge coupling to be sizeable, DM relic usually gets under-abundant due to large annihilation rate of DM into $Z'$ pairs. We consider a tiny kinetic mixing between $U(1)_D$ and $U(1)_Y$ of the SM by virtue of which DM can be produced from the SM bath via freeze-in mechanism \cite{Hall:2009bx, Bernal:2017kxu}. However, due to strong annihilation rate of DM into $Z'$ pairs, the final relic remains under-abundant. This requires some additional source of DM production to fill the deficit caused due to strong DM annihilations into $Z'$ pairs.

	In addition to DM, the SM also can not explain the origin of neutrino mass and mixing, as verified at neutrino oscillation experiments \cite{Zyla:2020zbs, Mohapatra:2005wg}. BSM framework must be invoked to explain non-zero neutrino mass as in the SM, there is no way to couple the left handed neutrinos to the Higgs field in the renormalisable Lagrangian due to the absence of right handed neutrinos. Conventional neutrino mass models based on seesaw mechanism \cite{Minkowski:1977sc, GellMann:1980vs, Mohapatra:1979ia, Schechter:1980gr, Mohapatra:1980yp, Lazarides:1980nt, Wetterich:1981bx, Schechter:1981cv, Foot:1988aq} typically involve introduction of heavy fields like right handed neutrinos (RHN).

	Motivated by these, we consider a self-interacting dark matter scenario where one of the RHNs, introduced for the purpose of generating light neutrino masses, can also play a non-trivial role in late time non-thermal production of DM, thus filling the deficit caused due to strong DM annihilations into $Z'$. This gives rise to RHN portal SIDM scenario while the RHN is connected to light neutrino masses. It should also be noted that neutrino portal DM have been studied in different contexts in earlier works like \cite{Falkowski:2009yz, Macias:2015cna, Batell:2017rol, Batell:2017cmf, Bandyopadhyay:2018qcv, Chianese:2018dsz, Blennow:2019fhy, Lamprea:2019qet, Chianese:2019epo, Bandyopadhyay:2020qpn, Berlin:2018ztp, Hall:2019rld,Biswas:2021kio}. However, in most of these works, either DM coupling directly with the SM lepton doublet was considered or a portal via heavy right handed or Dirac neutrinos responsible for DM freeze-out or freeze-in were discussed. In \cite{Biswas:2021kio}, DM-SM coupling via RHN was considered where RHNs are part of light Dirac neutrino with interesting cosmological consequences due to enhanced relativistic degrees of freedom $\Delta N_{\rm eff}$. In the present work, the RHNs can be produced thermally from SM bath with the lightest of them acquiring a thermal relic after freeze-out which then gets converted into SIDM relic due to its late time decay into the latter. After showing the generic features of this setup relevant for the required self-interactions and relic abundance of DM, we consider two specific seesaw models namely, gauged $B-L$ and scotogenic models both of which can lead to thermal production of the RHN and generate its required relic. We then constrain the model parameters from the rfequirement of generating correct relic abundance of the lightest RHN which gets converted into SIDM relic later. We also find that the models remain predictive at experiments like the large hadron collider (LHC) and experiments looking for charged lepton flavour violation, in addition to experiments related to astrophysical observations as well as DM direct detection.

	
	This paper is organized as follows. In section \ref{model}, we present the minimal setup of RHN portal SIDM showing the details of self-interaction, relic abundance and direct detection prospects. In section \ref{examples}, we outline two specific UV complete realizations discussing the connection to neutrino mass and other phenomenological prospects. We finally conclude in section \ref{sec:conclude}.			
	
	\section{SIDM via RHN portal}
	\label{model}
	In this section, we discuss the basic ingredients of the model and relevant phenomenology. The BSM particle content is shown in table \ref{tab:tab2} along with their transformations under the gauge symmetry. The SM gauge symmetry is extended by an Abelian $U(1)_D$ gauge symmetry under which an SM singlet Dirac fermion $\chi$ is charged. The right handed neutrino $N_R$ uncharged under $U(1)_D$ can act like a portal between the SM and dark sectors. A singlet scalar $\Phi$ with non-zero $U(1)_D$ charge not only lead to spontaneous breaking of $U(1)_D$, but also couples $N_R$ with DM fermion $\chi$ which plays a crucial role in generating observed DM relic.

	\begin{table}[h!]
			\begin{tabular}{|c|c|c|c|}
				\hline \multicolumn{2}{|c}{Fields}&  \multicolumn{1}{|c|}{ $ SU(3)_C \otimes SU(2)_L \otimes U(1)_Y $  $\otimes  U(1)_D $  } \\ \hline
				{Fermion} &  $N_R$&  ~~1 ~~~~~~~~~~~1~~~~~~~~~~0~~~~~~~~~~ 0 \\ [0.5em] 
				& $\chi$  & ~~1 ~~~~~~~~~~~1~~~~~~~~~~0~~~~~~~~~~ 1 \\
				[0.5em] \cline{1-3}
				{Scalars} & 
				$\Phi$ & ~~1 ~~~~~~~~~~~1~~~~~~~~~~0~~~~~~~~~~ -1 \\
				\hline
			\end{tabular}
			\caption{Charge assignment of BSM fields under the gauge group $G \equiv G_{\rm SM} \otimes U(1)_D$  where $G_{\rm SM}\equiv SU(3)_c \otimes SU(2)_L \otimes U(1)_Y$ .}
			\label{tab:tab2}
		\end{table}	
		
		Owing to the BSM particle content and their charge assignments as shown in table \ref{tab:tab2}, the generic Lagrangian with the interactions relevant for determining the DM abundance in the considered scenario is given by
		
		\begin{equation}
			\label{Lagrangian}
			\mathcal{L}_{DM} \supset i ~\overline{\chi}~ \gamma^\mu~ D_\mu~ \chi - m_{\chi} ~\overline{\chi} ~\chi -\frac{1}{2}M_{N_R} \overline{N^c_R} N_R-  y ~\chi ~\Phi~ N_R - y' \overline{\chi} \Phi^* N_R + \frac{\epsilon}{2}B^{\alpha \beta}Y_{\alpha\beta} 
		\end{equation}
		
		where $D_\mu = \partial_\mu + i g_D Z'_\mu$ and $B^{\alpha\beta}, Y_{\alpha \beta}$ are the field strength tensors of $U(1)_D, U(1)_Y$ respectively with $\epsilon$ being the kinetic mixing between them. {We consider the two Yukawa couplings between $\chi, N_R$ to be identical ($y = y'$) for simplicity and denote them by $y$ hereafter.} The Lagrangian involving singlet scalar $\Phi$ can be written as
		\begin{equation}
			\mathcal{L}_{\Phi} = (D_\mu \Phi)^\dagger (D^\mu \Phi) + m^2_\Phi \Phi^\dagger \Phi - \lambda_{\phi} (\Phi^{\dagger} \Phi)^2 -\lambda_{\Phi H} (\Phi^{\dagger} \Phi) (H^{\dagger} H)
			\label{scalarL}
		\end{equation}
		The singlet scalar can acquire a non-zero vacuum expectation value (VEV) $\langle \Phi \rangle =u$ giving rise to $U(1)_D$ gauge boson mass $M_{Z'}= g_D u$. The same singlet scalar VEV can also generate a mixing between $\chi$ and $N_R$ leading to pseudo-Dirac states. However, the corresponding Yukawa coupling $Y$ is very small as required from relic abundance criteria to be discussed below. Therefore, we continue to consider DM as pure Dirac states assuming $Y u \ll M_{N_R}$. While there can be multiple copies of RHNs, as required from neutrino mass requirement, we consider only the lightest of them $(N_1)$ to couple to DM for the sake of minimality. 
		

		\subsection{Dark Matter Self-interaction}
		\label{self_int}
		As discussed above, the DM is assumed to be a pure Dirac fermion having vector like couplings to $Z'$ of type $g' Z'_{\mu}\overline{\chi}\gamma^\mu\chi$. As a result
		the DM can have elastic self-scatterings mediated via $Z'$. In order to have sufficient self-interactions, one needs to satisfy the requirement on cross-section as $\sigma \sim 1~~cm^2 (\frac{m_\chi}{g}) \approx 2\times 10^{-24}~~cm^2 (\frac{m_\chi}{GeV})$, which is many orders of magnitude larger than the typical weak-scale cross-section ($\sigma \sim 10^{-36} cm^2$). Such large self-interacting cross-sections can naturally be realized with a light mediator or a dark boson much lighter compared to the typical weak scale mediators. For such light mediator which is same as $U(1)_D$ gauge boson $Z'$ in our setup, the self-interaction of non-relativistic DM can be described by a Yukawa type potential given as
		\begin{equation}
			V(r)= \pm \frac{\alpha_D}{r}e^{-M_{Z'}r}.
			\label{eqYuk}
		\end{equation} 
		In the above equation, the + (-) sign denotes repulsive (attractive) potential and $\alpha_D = g^2_D /4\pi$ is the dark fine structure constant. While $\chi \overline{\chi}$ interaction is attractive, $\chi \chi$ and $\overline{\chi} \overline{\chi}$ are repulsive. To capture the relevant physics of forward scattering divergence for the self-interaction we define the transfer cross-section $\sigma_T$ as~\cite{Feng:2009hw,Tulin:2013teo,Tulin:2017ara}
		\begin{equation}
			\sigma_T = \int d\Omega (1-\cos\theta) \frac{d\sigma}{d\Omega}.
		\end{equation}
		
		In the Born Limit ($\alpha_D m_\chi/M_{Z'}<< 1$), for both attractive as well as repulsive potentials, the transfer cross-section is
		\begin{equation}
			\sigma^{\rm Born}_T = \frac{8 \pi \alpha^2_D}{m^2_\chi v^4} \Bigg(\ln(1+ m^2_\chi v^2/M^2_{Z'} ) - \frac{m^2_\chi v^2}{M^2_{Z'}+ m^2_\chi v^2}\Bigg).
		\end{equation} 
		Outside the Born regime ($\alpha_D m_\chi /M_{Z'} \geq 1 $), we have two distinct regions {\it viz} the classical region and the resonance region. In the classical limit ($m_\chi v/M_{Z'}\geq 1$), the solutions for an attractive potential is given by\cite{Tulin:2013teo,Tulin:2012wi,Khrapak:2003kjw}
		\begin{equation}
			\sigma^{\rm classical}_T (\rm Attractive)=\left\{
			\begin{array}{l}
				
				\frac{4\pi}{M^2_{Z'}}\beta^2 \ln(1+\beta^{-1}) ~~~~~~~~~~~~~~\beta \leqslant 10^{-1}\\
				\frac{8\pi}{M^2_{Z'}}\beta^2/(1+1.5\beta^{1.65}) ~~~~~~~~~~~10^{-1}\leq \beta \leqslant 10^{3}\\
				\frac{\pi}{M^2_{Z'}}( \ln\beta + 1 -\frac{1}{2}ln^{-1}\beta) ~~~~~~~~\beta \geq 10^{3}\\
			\end{array}
			\right.
		\end{equation}  
		
		\begin{equation}
			\sigma^{\rm classical}_T (\rm Repulsive)=\left\{
			\begin{array}{l}
				
				\frac{2\pi}{M^2_{Z'}}\beta^2 \ln(1+\beta^{-2}) ~~~~~~~~~~~~~~\beta \leqslant 1\\
				\frac{\pi}{M^2_{Z'}}(\ln 2\beta^2 -\ln ~\ln 2\beta)^2 ~~~~~~~~~~~ \beta \geq 1\\
			\end{array}
			\right.
		\end{equation}   
		where $\beta = 2\alpha_D M_{Z'}/(m_\chi v^2)$.
		\begin{figure}[h!]
			\includegraphics[scale=0.4]{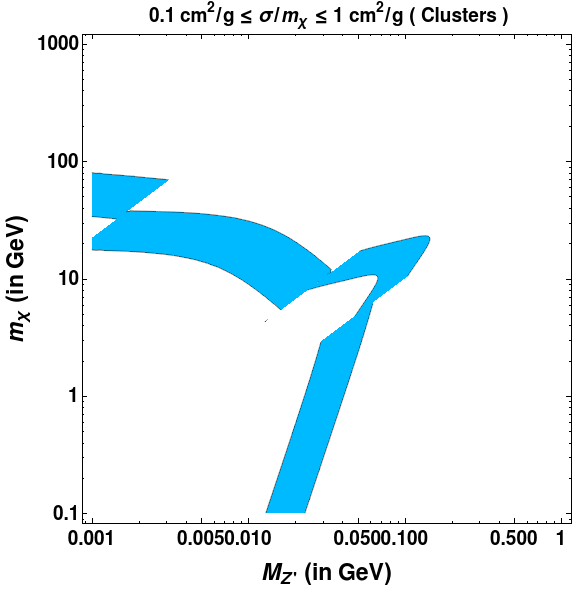}
			\hfil
			\includegraphics[scale=0.4]{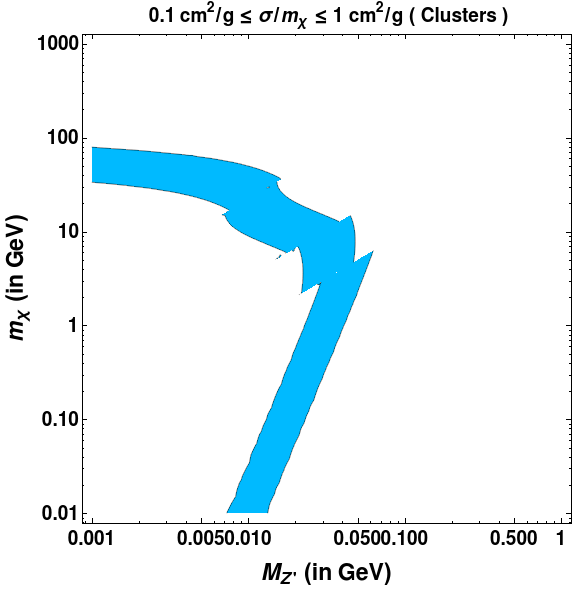}
			\caption{[Left]: Attractive;  [Right]: Repulsive self-interaction cross-section in the range $0.1-1 \; {\rm cm}^2/{\rm g}$ for clusters ($v\sim1000 \; {\rm km/s}$). Sky blue colour  represents regions of parameter space where $0.1 \; {\rm cm}^2/{\rm g} < \sigma/m_\chi <1 \;{\rm cm}^2/{\rm g}$.}
			\label{sidm1}
		\end{figure}
		\begin{figure}[h!]	
			\includegraphics[scale=0.4]{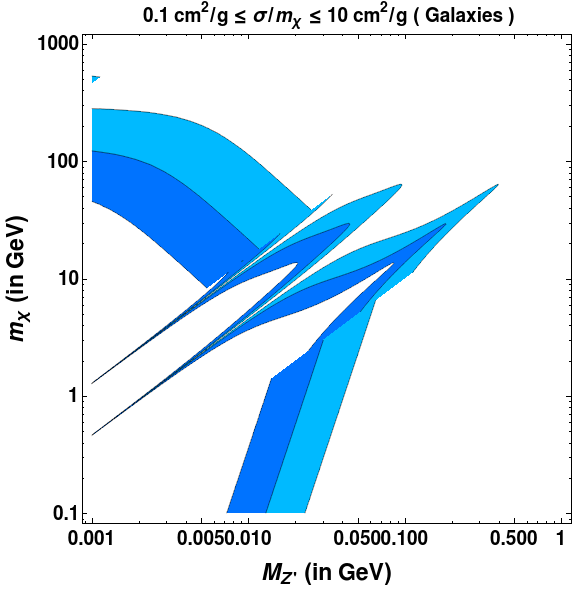}
			\hfil
			\includegraphics[scale=0.4]{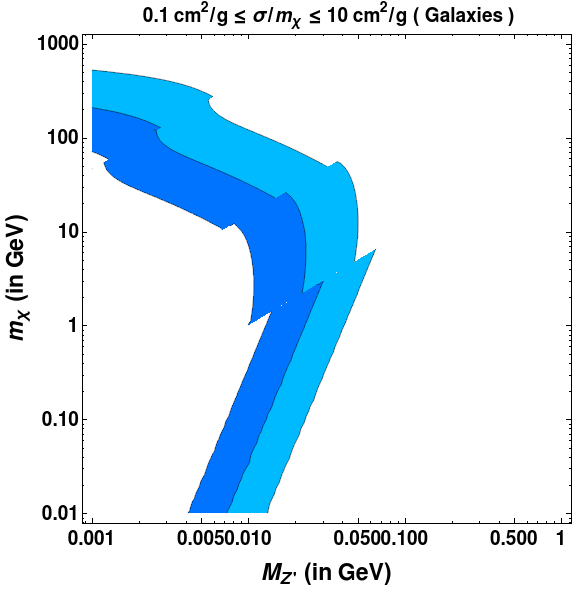}
			\caption{[Left]: Attractive;  [Right]: Repulsive self-interaction cross-section in the range  $0.1-10 \; {\rm cm}^2/{\rm g}$ for galaxies ($v\sim200 \; {\rm km/s}$). Blue colour represents regions of parameter space where $1 \; {\rm cm}^2/{\rm g} < \sigma/m_\chi <10 \; {\rm cm}^2/{\rm g}$; Sky blue colour  represents regions of parameter space where $0.1 \; {\rm cm}^2/{\rm g} < \sigma/m_\chi <1 \;{\rm cm}^2/{\rm g}$.}
			\label{sidm2}
		\end{figure}
		\begin{figure}[h!]	
			\includegraphics[scale=0.4]{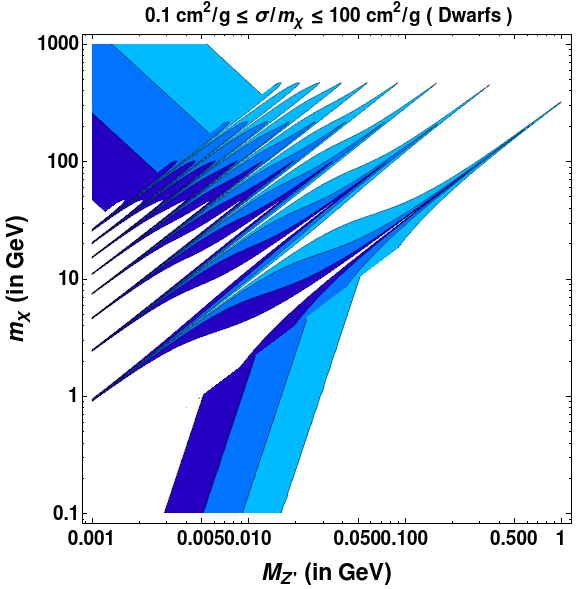}
			\hfil
			\includegraphics[scale=0.4]{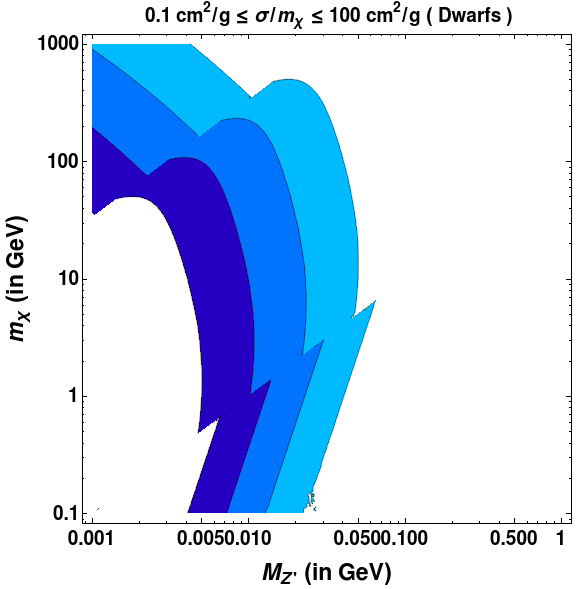}
			\caption{[Left]: Attractive;  [Right]: Repulsive self-interaction cross-section $0.1-100 \; {\rm cm}^2/{\rm g}$ for dwarfs ($v\sim10 \; {\rm km/s}$). Dark blue colour represents regions of parameter space where $10 \; {\rm cm}^2/{\rm g} < \sigma/m_\chi < 100 \; {\rm cm}^2/{\rm g}$ ;  Blue colour represents regions of parameter space where $1 \; {\rm cm}^2/{\rm g} < \sigma/m_\chi <10 \; {\rm cm}^2/{\rm g}$; Sky blue colour  represents regions of parameter space where $0.1 \; {\rm cm}^2/{\rm g} < \sigma/m_\chi <1 \;{\rm cm}^2/{\rm g}$.}
			\label{sidm3}
		\end{figure}
		
		Both the Born and the classical regimes do not provide us the mild velocity dependence in the cross-section required to explain the small-scale issues of CDM. However one interesting regime lies outside the classical regime for $\alpha_D m_\chi/M_{Z'} \geq 1, m_\chi v/M_{Z'} \leq 1$. This is the resonant regime, where quantum mechanical resonances and anti-resonances appear in $\sigma_T$ corresponding to (quasi-)bound states in the potential. In this regime, an analytical formula for $\sigma_T$ does not exist, and one has to solve the non-relativistic Schrodinger equation by partial wave analysis. Here we use the non-perturbative results for s-wave (l=0) scattering within the resonant regime obtained by approximating the Yukawa potential to be a Hulthen potential $\Big( V(r) = \pm \frac{\alpha_D \delta e^{-\delta r}}{1-e^{-\delta r}}\Big)$, which is given by~\cite{Tulin:2013teo}
		
		\begin{equation}
			\sigma^{\rm Hulthen}_T = \frac{16 \pi \sin^2\delta_0}{m^2_\chi v^2}
		\end{equation}
		
		where l=0 phase shift is given in terms of the $\Gamma$ functions by
		\begin{equation}
			\delta_0 ={\rm arg} \Bigg(\frac{i\Gamma \Big(\frac{i m_\chi v}{k M_{Z'}}\Big)}{\Gamma (\lambda_+)\Gamma (\lambda_-)}\Bigg),~~~~ \lambda_{\pm} = \left\{
			\begin{array}{l}
				1+ \frac{i m_\chi v}{2 k M_{Z'}}  \pm \sqrt{\frac{\alpha_D m_\chi}{k M_{Z'}} - \frac{ m^2_\chi v^2}{4 k^2 M^2_{Z'}}} ~~~~ {\rm Attractive}\\
				1+ \frac{i m_\chi v}{2 k M_{Z'}}  \pm i\sqrt{\frac{\alpha_D m_\chi}{k M_{Z'}} + \frac{ m^2_\chi v^2}{4 k^2 M^2_{Z'}}} ~~~~ {\rm Repulsive}\\
			\end{array}
			\right.
		\end{equation}   
		with $k \approx 1.6$ being a dimensionless number. 
		
		We use these expressions for DM self-interaction cross-sections considering the astrophysical bounds on required self-interactions namely, $\sigma/m_\chi$ at different scales to constrain the model parameters. Keeping the dark gauge coupling fixed as $g_D=0.1$, in Fig.~\ref{sidm1},\ref{sidm2},\ref{sidm3}, we show the allowed parameter space in $m_\chi$ versus $M_{Z'}$ plane which gives rise to the required DM self-interaction cross-section ($\sigma/m_\chi$) in the range $0.1-1~{\rm cm}^2/{\rm g}$ for clusters ($v\sim1000~ \rm km/s$), $\sigma \in 0.1-10~{\rm cm}^2/{\rm g}$ for galaxies ($v\sim 200~ \rm km/s$) and $\sigma \in 0.1-100~{\rm cm}^2/{\rm g}$ for dwarf galaxies ($v\sim 10~ \rm km/s$) respectively. Since the mediator is a vector boson, there exists both attractive and repulsive self-interactions as opposed to the case with scalar mediator giving only attractive self-interactions. The sharp spikes seen in the left panel plots of Fig.~\ref{sidm2},\ref{sidm3} denote the patterns of quantum mechanical resonances and anti-resonances for the attractive potential. For a repulsive potential case, such resonances are absent as shown on the right panels of Fig.~\ref{sidm2},\ref{sidm3}. The resonant regime covers a large portion of the parameter space in the $m_\chi$ versus $M_{Z'}$ plane. For galactic and dwarf galactic scales with relatively small DM velocities, such features are more dominant. This is due to the fact that for a fixed $\alpha_D$, the condition $m_{\chi} v/M_{Z'} < 1$ dictates the onset of non-perturbative quantum mechanical effects. One can notice from these figures that even after using the constraints from required self-interactions, a wide range of DM mass remains allowed even though the mediator mass gets constrained within one or two orders of magnitudes except in the resonance regimes. While discussing other relevant DM phenomenology in upcoming sections, we will compare these allowed regions of parameter space from required self-interactions with the ones obtained using other relevant constraints.

\begin{figure}
			$$
			\includegraphics[scale=0.35]{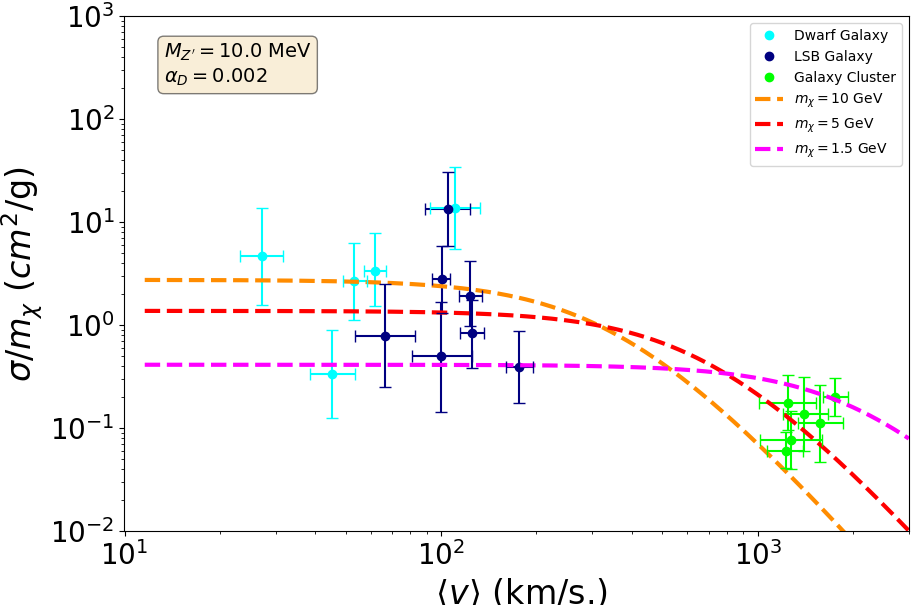}
			$$
			\caption{The self-interaction cross section per unit mass of DM as a function of average collision velocity.}
			\label{astrofit}
		\end{figure}

		Finally, we show the self-interaction cross-section per unit DM mass as a function of average collision velocity in Fig.\ref{astrofit}. The data points with error bars include observations of dwarfs (cyan), low surface brightness (LSB) galaxies (blue) and clusters (green)\cite{Kaplinghat:2015aga,Kamada:2020buc}. The dashed line correspond to different DM masses while other parameters are fixed as $M_{Z'}=10~\rm MeV$ and $\alpha_D = 0.002$. It is clear from Fig.\ref{astrofit} that the SIDM model we adopt can explain astrophysical observations of velocity dependent DM self-interaction appreciably well. It is worth mentioning here that, for fexed DM mass if we increase $Z'$ mass, the DM self-interaction cross-section gradually decreases and may not satisfy the astrophysical observation for the DM mass range we are interested in ({\it i.e.}$1-10$ GeV). One can, however, increase the corresponding gauge coupling to compensate for this decrease, as long as we are within perturbative limits.


		\subsection{Production of Dark Matter}
		\label{dm_production}
		
		Several production regimes for self-interacting DM exist in the literature~\cite{Kouvaris:2014uoa, Bernal:2015ova, Kainulainen:2015sva, Hambye:2019tjt, Cirelli:2016rnw, Kahlhoefer:2017umn, Belanger:2011ww}. While DM can have large enough self-interaction mediated by the light $Z'$ with sizeable $g_D$ sufficient to address the small-scale issues, it can interact with the thermal bath only via kinetic mixing of neutral vector bosons. Even though DM has coupling with RHN with the latter being assumed to be in thermal bath, the corresponding coupling is required to be small from late freeze-in criteria to be discussed below.

\begin{figure}
			\centering
			\includegraphics[scale=0.5]{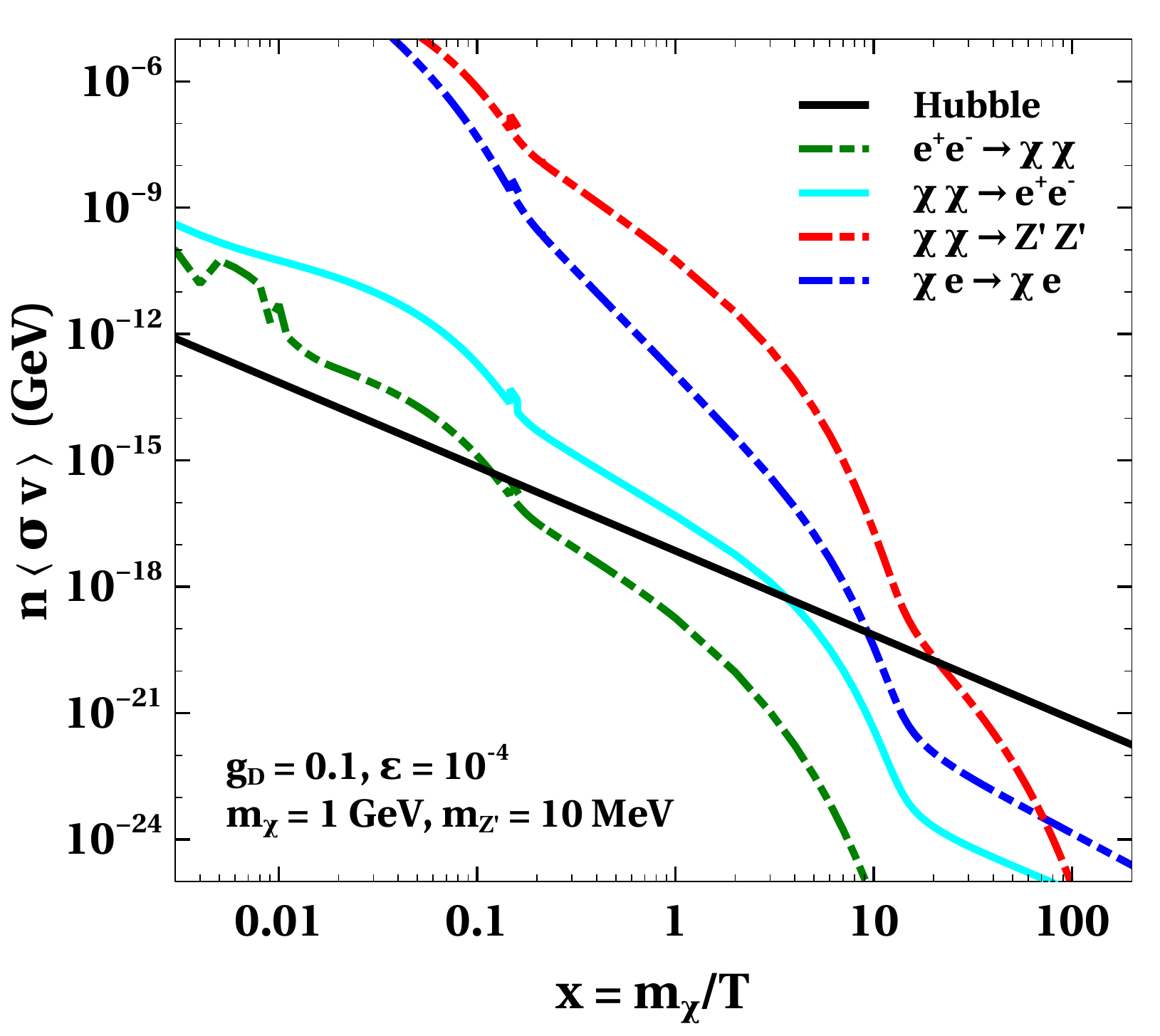}
			\caption{Comparison of different scattering processes involving DM with Hubble rate of expansion in radiation dominated era for kinetic mixing parameter $\epsilon = 10^{-4}$.}
			\label{decoupling4}
		\end{figure}

\begin{figure}[ht]
			\includegraphics[scale=0.3]{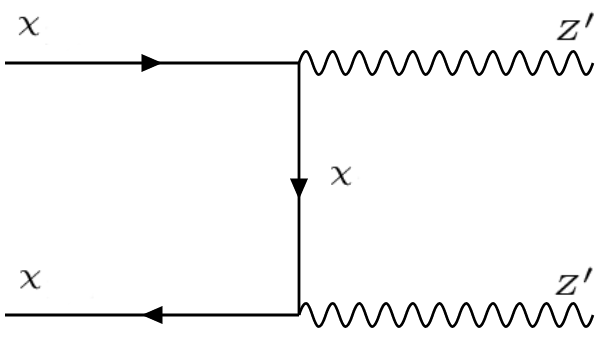} 
			\hfil
			\includegraphics[scale=0.3]{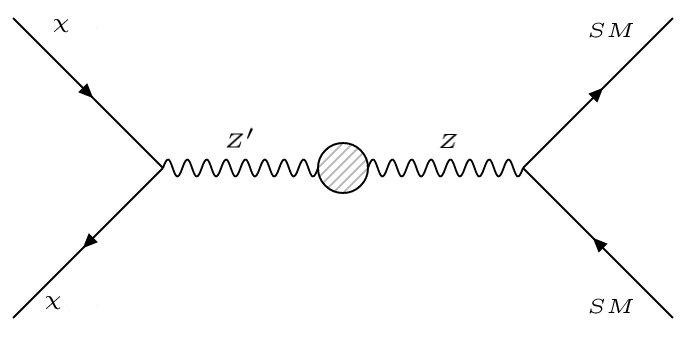}
			\caption{Feynman diagrams for dominant number changing processes of DM.}
			\label{feyn}
		\end{figure}

		For sizeable kinetic mixing ($\epsilon \geq 10^{-5}$), the DM can be in thermal equilibrium with the SM bath in the early universe. In Fig.~\ref{decoupling4}, we show the comparison of the rates of different processes considering $\epsilon = 10^{-4}$ with the Hubble expansion rate of the universe in a radiation dominated era.  For numerical analysis, the model has been implemented in standard packages like \texttt{LanHEP} \cite{Semenov:2014rea} and \texttt{CalcHEP} \cite{Belyaev:2012qa}. As we can see from Fig.~\ref{decoupling4}, all relevant processes are well above the Hubble expansion rate at early epochs keeping DM $\chi$ in thermal equilibrium. However, to get velocity dependent self-scattering we are considering heavier DM compared to the mediator {\it i.e.,} $m_{\chi} > M_{Z'}$ and therefore, DM has large annihilation cross section to $Z'$ pairs compared to its annihilation rates into SM particles. This can significantly lead to suppressed thermal relic of DM as we discuss below. The dominant number changing processes contributing to its thermal freeze-out are shown in Fig.~\ref{feyn}. 
		
\begin{figure}
			\centering
			\includegraphics[scale=0.5]{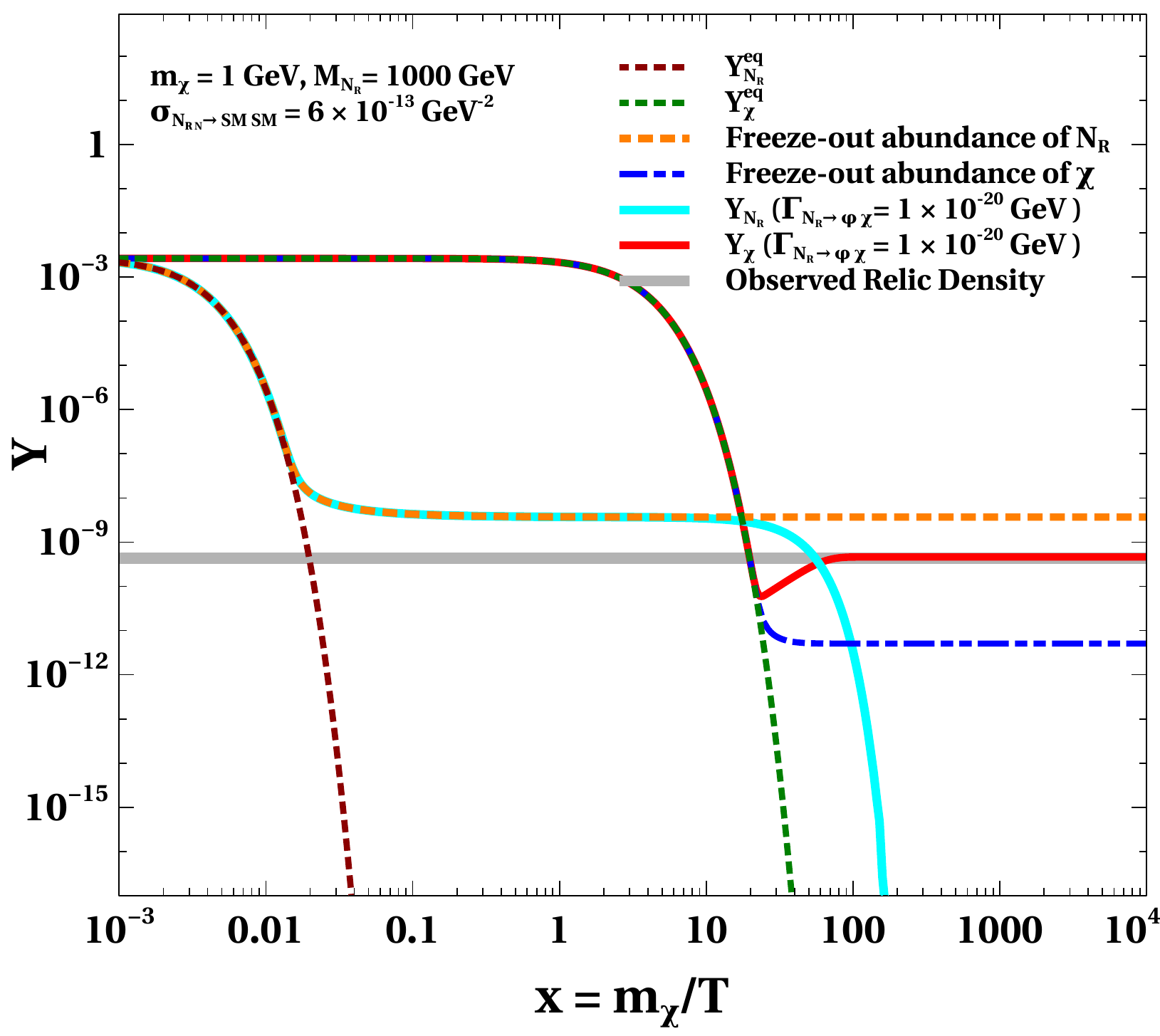}
			\caption{Comoving number densities of thermal dark sector particles considering different sub-processes indicated in the legends.}
			\label{Y_thermal}
		\end{figure}

		While the right panel diagram of Fig.~\ref{feyn} is suppressed by kinetic mixing, the left panel diagram (for sizeable gauge coupling $g_D \sim 0.1$ and GeV scale DM mass) leads to a cross-section which is much larger than the one governing typical WIMP annihilation. The thermally averaged cross-section for this process can be approximately given by
		\begin{equation}
			\langle\sigma v\rangle \sim \frac{\pi \alpha^2_D}{m^2_{\chi}}
		\end{equation}
		where $\alpha_D=g^2_D/(4\pi) $. As a consequence, the thermal freeze-out abundance is well below the correct ballpark given by Eq.~\eqref{dm_relic}. This is where the importance of the RHN portal arises as the RHN can decay at late time to DM bringing the relic back to its correct ballpark. To calculate the relic density precisely, we define comoving number densities of these particles as $Y_{\chi}=n_{\chi}/s(T)), Y_{N_{R}} = n_{N_{R}}/s(T)$ with $s(T)$ being the entropy density and write down the coupled Boltzmann equations for the DM $\chi$ and the right handed neutrino $N_{R}$ as follows. 
		
		\begin{equation}
			\footnotesize{
				\begin{aligned}
					&\frac{dY_{N_R}}{dx}= -\frac{s(m_{\chi})}{x^2  H(m_{\chi})} \langle\sigma v \rangle_{\rm F.O.}^{N_R} (Y^2_{N_R} -\big(Y^{\rm eq}_{N_R}\big)^2) -\frac{x}{H(m_{\chi})}\langle \Gamma_{N_R \rightarrow \phi \chi}\rangle Y_{N_R} \,\, ,
					\\&
					\frac{dY_{\chi}}{dx}=-\frac{s(m_{\chi})}{x^2  H(m_{\chi})}  \langle\sigma(\chi \chi  \to Z' Z') v\rangle (Y^2_{\chi}-\big(Y^{\rm eq}_{\chi}\big)^2))
					+ \frac{x}{H(m_{\chi})} \langle \Gamma_{N_R\rightarrow \phi \chi}\rangle Y_{N_R}\,\, ,
			\end{aligned}}
			\label{eq:BE1}
		\end{equation} 
		where $x=m_{\chi}/T$. The evolution of comoving number densities of these particles for a typical set of parameters: ( $m_\chi = 1 {~\rm GeV}, M_{N_{R}}=1 {~\rm TeV}, M_{Z'}=10 ~{\rm MeV}, g_D=0.1$) is shown in Fig.~\ref{Y_thermal} by solving the coupled Boltzmann equations given by Eq.~\eqref{eq:BE1}. While we have not specified the interactions of $N_R$ with SM particles in this minimal setup, we assume $N_R$ to be in equilibrium with the thermal bath initially and thereafter freezes out with freeze-out cross-section $\langle \sigma v \rangle _{\rm F.O.}^{N_R}= 6\times 10^{-13} \, {\rm GeV}^{-2}$. The equilibrium number densities of $\chi$ and $N_R$ are shown by dotted green and dotted brown curves respectively in Fig.~\ref{Y_thermal}. The thermal freeze-out of the $\chi$ is depicted by the dashed blue curve and that of RHN is depicted by the dashed yellow curves. The thermal freeze-out followed by the decay of $N_{R}$ is depicted by the dashed cyan coloured curve, which lifts the DM relic to the observed value as depicted by the red graph. Similar discussions on DM having deficit in its thermal abundance, followed by late decay of another particle can be found in earlier works \cite{Molinaro:2014lfa, Borah:2017dfn, Biswas:2018ybc, Dutta:2021wbn, Borah:2021jzu, Borah:2021yek, DEramo:2018khz} where a hybrid of freeze-out and freeze-in mechanisms were utilized to get correct DM relic abundance. Use of such hybrid setup in other contexts may be found in \cite{Feng:2003uy, Borah:2019bdi, Borah:2018gjk} and references therein.

\begin{figure}
			\centering
			\includegraphics[scale=0.5]{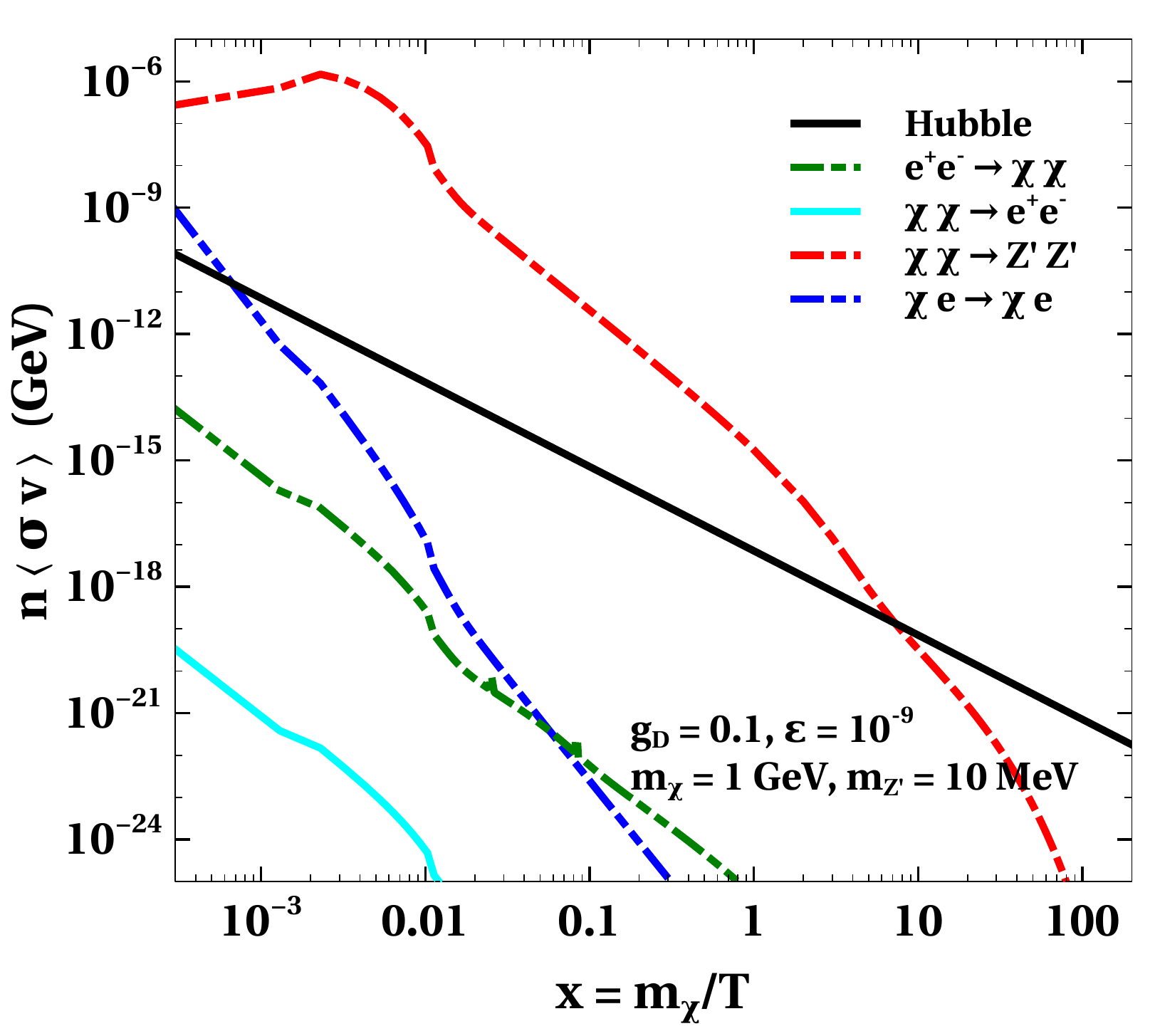}
			\caption{Comparison of different scattering processes involving DM with Hubble rate of expansion in radiation dominated universe for kinetic mixing parameter $\epsilon = 10^{-9}$.}
			\label{Decoupling8}
		\end{figure}

		However it is to be noted such large kinetic mixing can be in tension with DM direct search bounds from  CRESST-III~\cite{Abdelhameed:2019hmk}, XENON1T \cite{Aprile:2018dbl} depending upon DM as well as mediator masses and gauge coupling. Also, for light $Z'$ in sub-GeV scale, dark photon searches can put such large kinetic mixing in tension with low energy data \cite{Bauer:2018onh}. To be consistent with such bounds, we consider the kinetic mixing to be very small $(\sim 10^{-8})$, while simultaneously satisfying the large DM self-interaction criteria governed by sizeable $g_D$ and light $Z'$. But if the kinetic mixing parameter is very small, the DM is not necessarily in thermal equilibrium in the early universe. To check whether DM-SM interactions for such small kinetic mixing can indeed reach equilibrium in the early universe or not, we compare the rates of different annihilation processes with the Hubble expansion rate of the radiation dominated universe. The comparison is shown in Fig.~\ref{Decoupling8} considering kinetic mixing parameter $\epsilon = 10^{-9}$.
		
		From Fig.~\ref{Decoupling8}, it is evident that $\chi \chi \to Z'Z'$ interaction remains in equilibrium until very late epoch $x \sim 10$, while DM-SM number changing interactions mostly remain out of equilibrium throughout. However, the number conserving scattering process ${\rm DM} \; e \rightarrow {\rm DM} \; e$, which is responsible for keeping both the sectors in kinetic equilibrium, decouples around $x \sim 0.0007$. So prior to $x \sim 0.0007$, the dark sector temperature (denoted by $T'$) is same as that of thermal bath (denoted by $T$).  After $x \sim 0.0007$, the temperature of the dark sector evolves independently of the thermal bath until $x \sim 100$ when all the dark sector particles including light vector boson $Z'$ become non-relativistic and no longer contribute to the entropy degrees of freedom. Between $x \sim 0.0007$ and $x \sim 100$, the ratio of the visible and dark sector temperatures can be obtained by conserving the total entropy density separately in the two sectors. Considering the kinetic decoupling temperature to be $T_D$, we can relate the temperature of the two sectors as : 
		\begin{equation}
			\frac{T'}{T} = \left( \frac{g^{\rm SM}_{*s} (T)}{g^{\rm SM}_{*s} (T_D)} \right)^{1/3}.
			\label{eqdec}
		\end{equation}
		Here $g^{\rm SM}_{*s} (T)$ is the relativistic entropy degrees of freedom in the SM which goes into the calculation of relativistic entropy density $s(T)=\frac{2\pi^2}{45}g_{*s}(T)T^3$. Since the above relation~\eqref{eqdec} between two temperatures is valid for $T<T_D$, we naturally have $g^{\rm SM}_{*s} (T) < g^{\rm SM}_{*s} (T_D)$ leading to $T'< T$. This is also understood from the fact that SM bath temperature receives additional entropy contributions from the species which keep getting decoupled gradually. Within the decoupled dark sector itself, the DM particles can transfer their entropy into much lighter $Z'$ bosons once $T'$ falls below DM mass. This corresponds to an enhancement of dark sector temperature for $T'< m_{\rm DM}=m_{\chi}$ by $(13/6)^{1/3}$, a factor close to unity. We have ignored this additional but tiny enhancement in the numerical calculations to be discussed below. For dark sector temperature, we can similarly define a dimensionless integration variable $x'$, related to the usual variable $x$ as 
		\begin{equation}
			x'=\frac{m_{\rm \chi}}{T'}=\left (\frac{T}{T'}\right ) x.
		\end{equation}
		
		In the case of small kinetic mixing, we assume negligible initial density for the DM, and the DM can be produced from the bath particles by freeze-in mechanism \cite{Hall:2009bx}. However, due to strong $\chi \chi \to Z'Z'$ cross-section, such non-thermally produced DM will quickly annihilate into light $Z'$ boson pairs depleting the relic further. The right handed neutrino $N_R$ can decay to scalar $\phi$ and the DM $\chi$ at late epoch to bring the relic to the correct ballpark. Unlike the DM whose interactions with the SM bath are suppressed due to tiny kinetic mixing, the right handed neutrino $N_R$ can be in thermal equilibrium with the SM leading to its thermal freeze-out followed by its late decay into DM.
		
		As DM and $N_R$ masses can be very different from each other, the freeze-out abundance of $N_R$ need not necessarily match the required relic density of the DM always. However, both the thermal freeze-out cross-section of $N_R (\langle\sigma v \rangle^{^{N_R}}_{_{ \rm F.O.}})$ and the decay width of $N_R$ ($\Gamma_{N_{R} \to \phi \chi})$ play crucial roles in determining the the correct relic of the DM. In fact, there exists strong correlation between $\langle\sigma v \rangle^{^{N_R}}_{_{ \rm F.O.}}$ and $\Gamma_{N_{R} \to \phi \chi}$ from the requirement of producing correct relic of DM of a particular mass, which we discuss below.
		
		To calculate the relic density for freeze-in case with small kinetic mixing, we again define comoving number densities as $Y_{\chi}=n_{\chi}/s'(T'(T)), Y_{N_{R}} = n_{N_{R}}/s(T)$ and write down the coupled Boltzmann equations for DM $\chi$ and the right handed neutrino $N_{R}$ as follows. 
		.
		\begin{equation}
			\footnotesize{
				\begin{aligned}
					&\frac{dY_{N_R}}{dx}= -\frac{s(m_{\chi})}{x^2  H(m_{\chi})\Big( \frac{T'}{T}\Big)} \langle\sigma v \rangle^{^{N_R}}_{_{ \rm F.O.}} (Y^2_{N_R} -\big(Y^{\rm eq}_{N_R}\big)^2) -\left ( \frac{T'}{T}\right )^2\frac{x}{H(m_{\chi})}\langle \Gamma_{N_R \rightarrow \phi \chi}\rangle Y_{N_R} \,\, ,
					\\&
					\frac{dY_{\chi}}{dx}=\Bigg( \frac{T'}{T}\Bigg)^2\Bigg[\frac{s(m_{\chi})}{x^2  H(m_{\chi})} \Big(\langle\sigma(e^+ e^- \to \chi \chi) v\rangle (Y^{\rm eq}_{\chi})^2  - \langle\sigma(\chi\chi  \to Z' Z') v\rangle Y^2_{\chi} \Big)
					+ \frac{x\Big(\frac{g_{*s}(T_D)}{g'_{*s}(T_D)}\Big)}{H(m_{\chi})} \langle \Gamma_{N_R\rightarrow \phi \chi}\rangle Y_{N_R}\Bigg]\,\, ,
			\end{aligned}}
			\label{eq:BE}
		\end{equation}
		where $x=\frac{m_{\chi}}{T}$, $s(m_{\chi})= \frac{2\pi^2}{45}g_{*S}m^3_{\chi}$ , $H(m_{\chi})=1.67 g^{1/2}_*\frac{m^2_{\chi}}{M_{\rm Pl}}$ and $\langle \sigma v \rangle_{F.O.}^{N_R}$ represents the thermally averaged freeze-out cross-section~\cite{Gondolo:1990dk} of $N_R$. Also, $\langle \Gamma_{N_R\rightarrow \phi \chi}\rangle$ represents the thermally averaged decay width of the process $N_R \rightarrow \phi \chi$. The relevant cross-section and decay widths are given in Appendix~\ref{appen1}. Note that for sufficiently large kinetic mixing, the process $e^+ e^- \to \chi \chi$ becomes both-directional and $T'=T$, {\it i.e., } DM comes to thermal equilibrium and we get back to Eq.~\eqref{eq:BE1}.
		
		To solve the coupled Boltzmann equations given by Eq.~\eqref{eq:BE}, we divide the integration range into three regions as follows.
		
		\begin{itemize}
			\item i) $x < 0.0007$, where both the dark and the visible sectors share the same temperature $T=T'$ (see Fig.~\ref{Decoupling8}).
			\item ii) $0.0007<x<100$ where the dark sector is decoupled from the thermal bath and its temperature evolves according to Eq.~\eqref{eqdec}. 
			\item iii) $x>100$, the dark sector ceases to contribute to relativistic entropy density.
		\end{itemize}
		
\begin{figure}[h!]
			\centering
			\includegraphics[height=7cm,width=7cm]{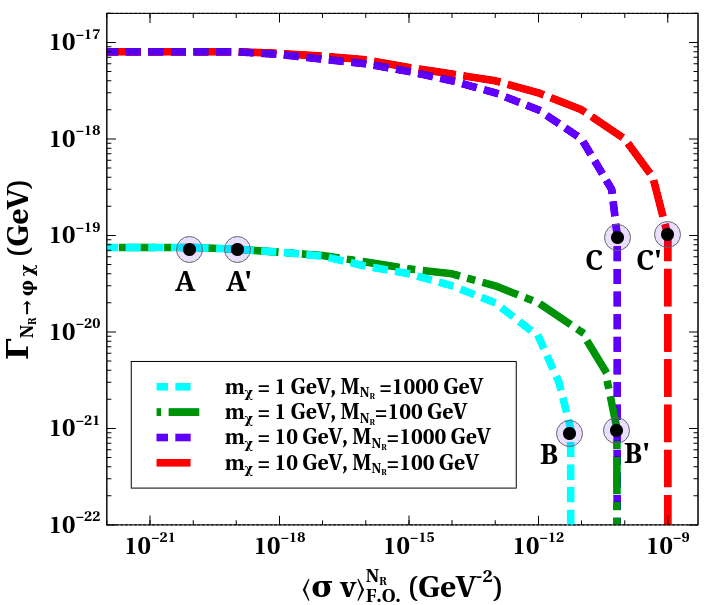}
			\includegraphics[height=7cm,width=7cm]{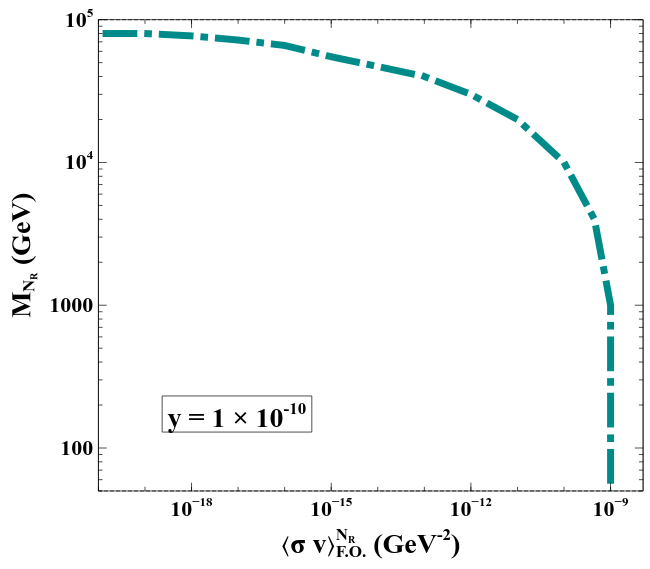}
			\caption{[Left]: Contours of correct DM relic in the plane of $\langle\sigma v \rangle^{^{N_R}}_{_{ \rm F.O.}}$ versus $\Gamma_{N_R \to \phi \chi}$. [Right]: Range of RHN mass $M_{N_R}$ for a fixed Yukawa coupling $y=1\times 10^{-10}$ giving the decay width in the correct range as shown in the left panel.}
	\label{relic_scan}
\end{figure}

		We solve the Boltzmann equations taking into account the temperature evolution in region i) and ii) mentioned above. For calculation of relic density, we consider the freeze-out cross-section of RHN $\langle\sigma v \rangle^{^{N_R}}_{_{ \rm F.O.}}$ and the decay width $\Gamma_{N_R \to \phi \chi}$ as free parameters and explain the details of how this cross-sections can be realized within the framework of a model in a later section~\ref{examples}. We find that there exists interesting correlation between $\langle\sigma v \rangle^{^{N_R}}_{_{ \rm F.O.}}$ and $\Gamma_{N_R \to \phi \chi}$ to give the correct relic for a DM of particular mass, which also depends upon the mass of $N_R$. We show the contours of correct relic for different cases in the plane of  $\langle\sigma v \rangle^{^{N_R}}_{_{ \rm F.O.}}$ versus $\Gamma_{N_R \to \phi \chi}$ in the left panel of Fig.~\ref{relic_scan}. We keep $Z'$ mass $M_{Z'}$=10 MeV, with DM and RHN masses as per the following benchmarks.
		
		\begin{itemize}
			\item i) $m_\chi = 1 \,{\rm GeV}, M_{N_R} = 1000 \,{\rm GeV}$ depicted by the thick dashed cyan curve.
			\item ii) $m_\chi = 1 \,{\rm GeV}, M_{N_R} = 100\, {\rm GeV}$ depicted by the thick dot-dashed green curve.
			\item iii) $m_\chi = 10\, {\rm GeV}, M_{N_R} = 1000\, {\rm GeV}$ depicted by the thick dashed blue curve.
			\item iv) $m_\chi = 10 \,{\rm GeV}, M_{N_R} = 100\, {\rm GeV}$ depicted by the thick  dashed red curve.
		\end{itemize}
		
To analyze the left panel of Fig.~\ref{relic_scan}, let us identify three distinct cases of DM production depending on  $\langle\sigma v \rangle^{^{N_R}}_{_{ \rm F.O.}}$ versus $\Gamma_{N_R \to \phi \chi}$. 

\begin{figure}[h!]
	\centering
	\includegraphics[scale=0.45]{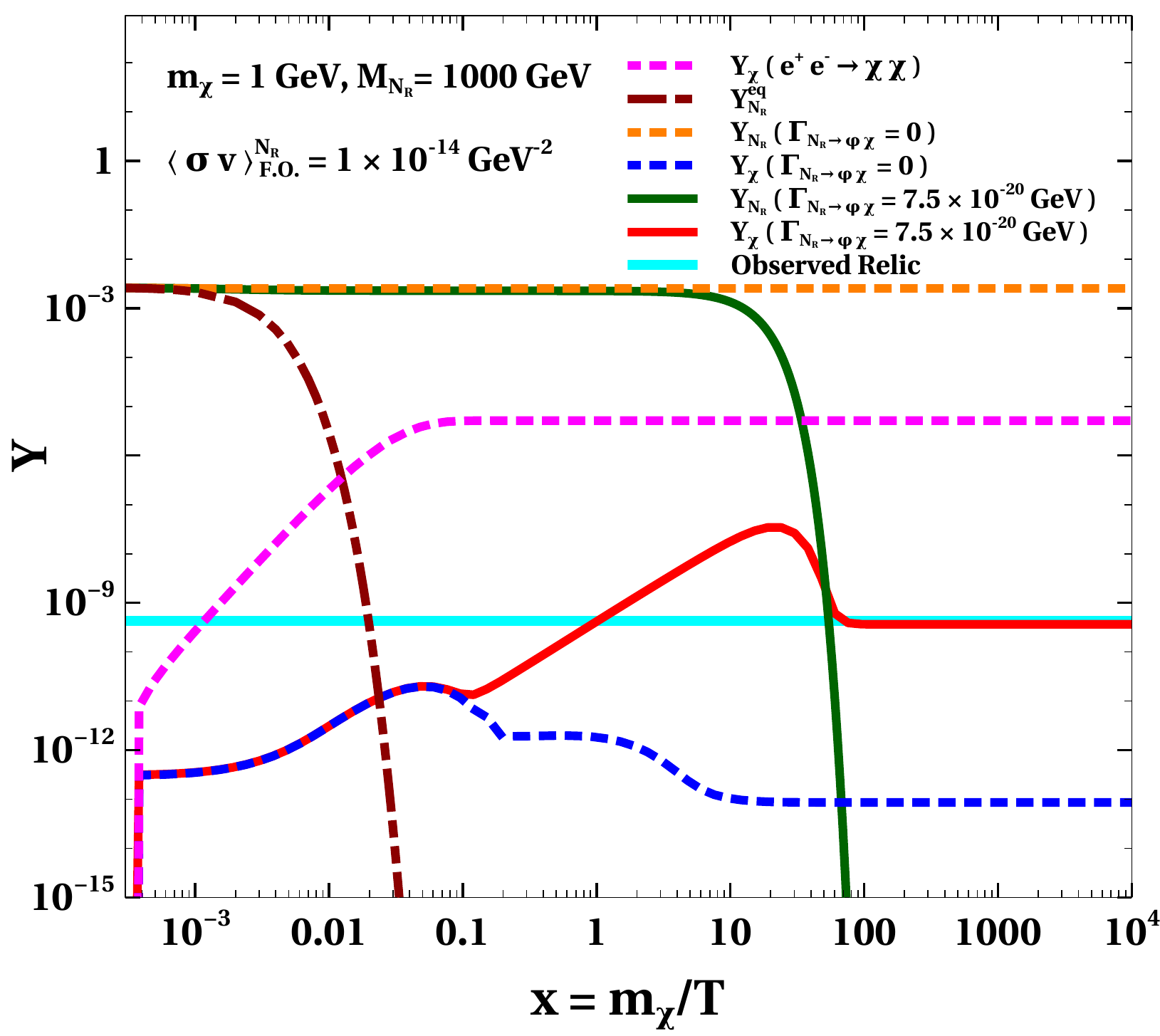}
	\includegraphics[scale=0.45]{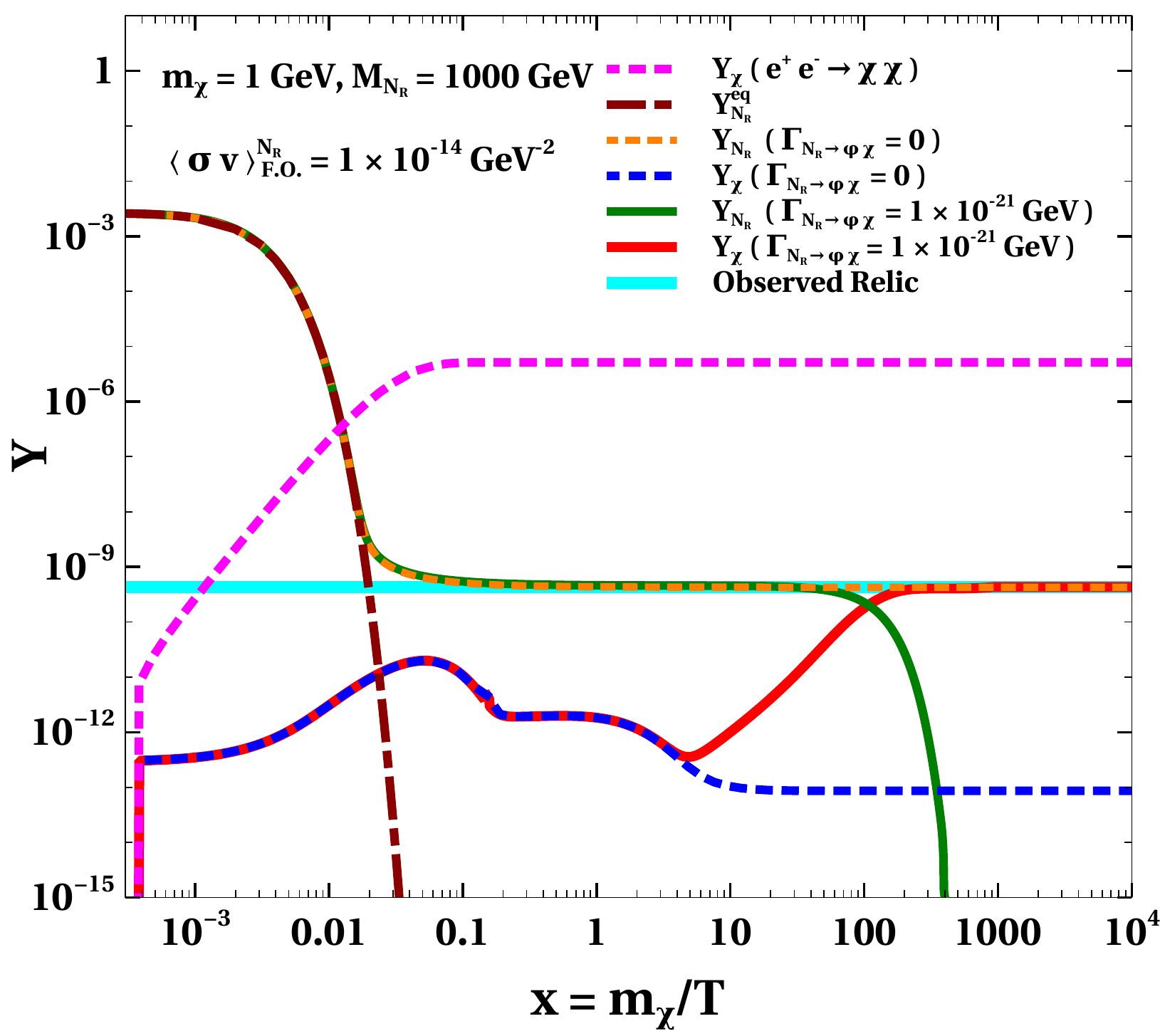}
	\caption{ (Left: Case-I, Right: Case-II.) Comoving number densities of dark sector particles considering different sub-processes indicated in the legends.}
	\label{relic1}
\end{figure}

{\bf Case-I:} When $\langle\sigma v \rangle^{^{N_R}}_{_{ \rm F.O.}}$ is very small, the freeze-out abundance of $N_R$ is large and can be comparable to the photon number density. If such a huge abundance is transferred to DM at later epochs, it may lead to overproduction of DM as well. Therefore, to get the correct relic density for the DM, decay of $N_R$ must occur at such an early epoch so that the $\chi \chi \to Z'Z'$ annihilation is still very active\footnote{Note that $\chi \chi \to Z'Z'$ is active till $x \sim 13$ as seen in Fig.~\ref{Decoupling8}.}. As a result the huge DM relic obtained from $N_R$ decay can still settle down to the observed DM relic due to subsequent annihilation into $Z'$ pairs. For example, the point marked as A ($\langle\sigma v \rangle^{^{N_R}}_{_{ \rm F.O.}}= 1\times10^{-20}\, {\rm GeV}^{-2}$) in the left panel of Fig.~\ref{relic_scan}, indicates a particular value of the $N_R$ freeze-out cross-section with $N_R$ mass 1000 GeV. If the cross-section of $N_R$ is 
further small, then the freeze-out abundance of $N_R$ can be as large as the photon number density. At and below this cross-section, the decay width of $N_R$ to give correct relic density must be $\Gamma_{N_R \to \phi \chi}= 7.5\times 10^{-20}\,{\rm GeV}$. For better understanding, we show the corresponding evolution of comoving number densities in the left panel of Fig.~\ref{relic1}, to be discussed later.

\begin{figure}[h!]
	\centering
	\includegraphics[scale=0.5]{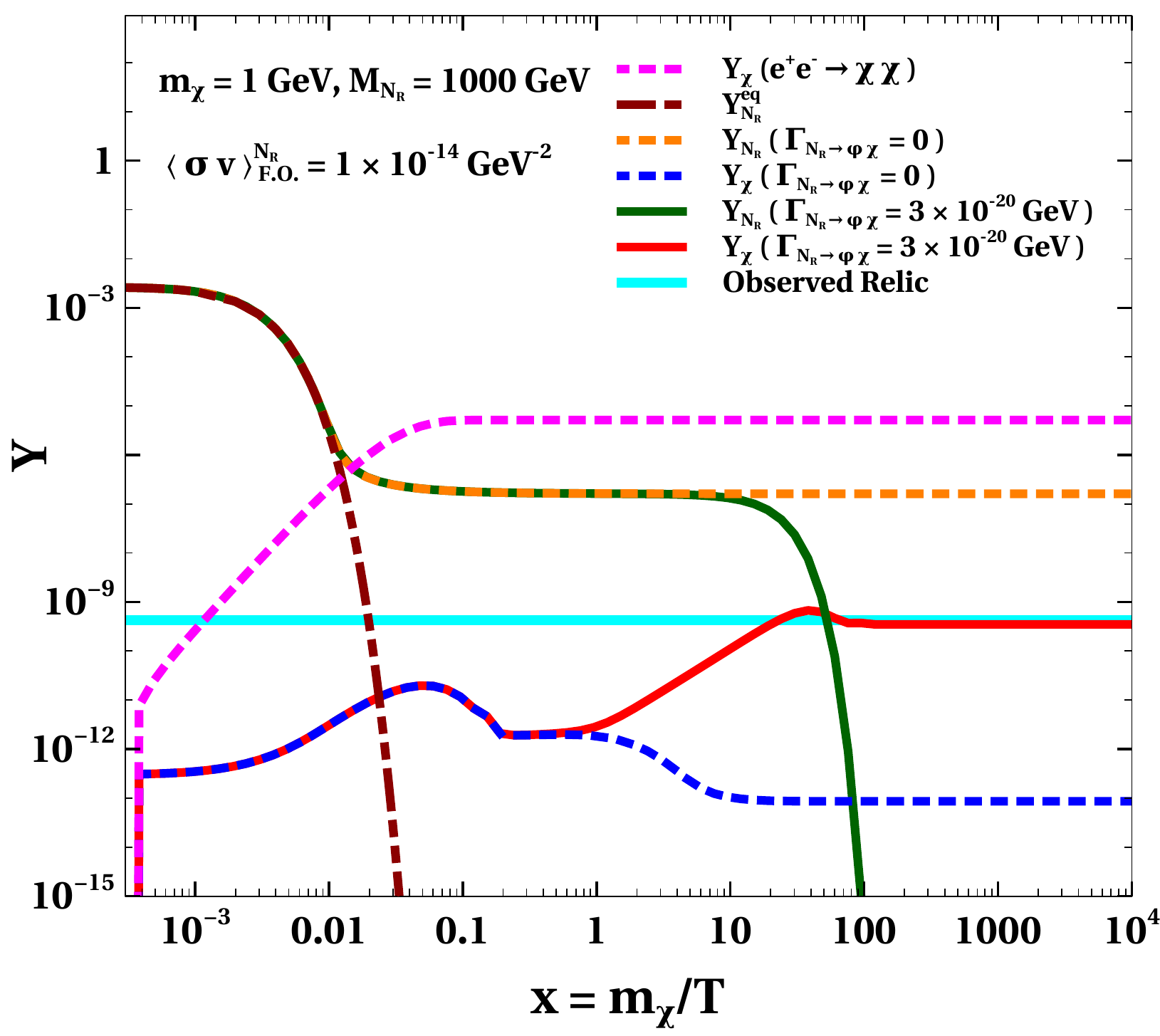}
	\caption{(Case-III) Comoving number densities of dark sector particles considering different sub-processes indicated in the legends.}
	\label{relic2}
\end{figure} 

{\bf Case-II:}  When $\langle\sigma v \rangle^{^{N_R}}_{_{ \rm F.O.}}$ is such that the freeze-out abundance of $N_R$ exactly matches the correct DM relic for a particular DM mass $m_\chi$, the decay of $N_R$ must occur at sufficiently late epochs\footnote{However, decay width can not be arbitrarily small as the decay must occur before the big bang nucleosynthesis (BBN) epoch in order to avoid extra entropy injection. Considering epoch of BBN to be around 1 sec, we get a conservative lower bound on the decay width as $\Gamma_{N_R \to \phi \chi}> 6.5 \times 10^{-25}{\rm GeV}$.} so that $\chi \chi \to Z'Z'$ remains almost ineffective when $N_R$ abundance gets converted into DM abundance. For example, the point marked as B in the left panel of Fig.~\ref{relic_scan} indicates such a scenario for $N_R=1000$ GeV and $m_\chi = 1$ GeV where the freeze-out abundance of $N_R$ due to its annihilation cross-section $\langle\sigma v \rangle^{^{N_R}}_{_{ \rm F.O.}}= 5.6 \times 10^{-12}\,{\rm GeV}^{-2}$ exactly matches the required relic for a 1 GeV DM ($4.4 \times 10^{-10}$ in terms of comoving number density). For this particular cross-section, the decay width $ \Gamma_{N_R \to \phi \chi} \leq 1\times 10^{-21}\,{\rm GeV}$ while satisfying lifetime bound from BBN limits. This upper bound on decay width of $N_R$ arises because for larger decay width and hence shorter lifetime, $N_R$ will convert to DM at early epochs when DM annihilation into $Z'$ pairs remain efficient further depleting its relic abundance. We show this explicitly in right panel of Fig~\ref{relic1}, to be discussed later.

{\bf Case-III:} If the $N_R$ freeze-out cross-section  $\langle\sigma v \rangle^{^{N_R}}_{_{ \rm F.O.}}$ has some intermediate values in between the ones mentioned in case-I and case-II above, correct relic can be obtained by a suitable combination of $\langle\sigma v \rangle^{^{N_R}}_{_{ \rm F.O.}}$ and $\Gamma_{N_R \to \phi \chi}$. As the cross-section gradually increases along x-axis in the left panel of Fig.~\ref{relic_scan}, decoupling of $N_R$ from the equilibrium is also delayed leading to decrease in freeze-out relic of $N_R$. Therefore, in order to get correct DM relic, decay of $N_R$ should also be delayed gradually (or $\Gamma_{N_R \to \phi \chi}$ should be smaller) so that $\chi \chi \to Z'Z'$ becomes less effective by the time the decay occurs. As an example, we show in Fig.~\ref{relic2}, the case of $\langle\sigma v \rangle^{^{N_R}}_{_{ \rm F.O.}}= 1 \times 10^{-14}\, {\rm GeV}^{-2}$ for $M_R = 1000$ GeV, which  is significantly higher than the one in case-I but lower than the one in case-II discussed above. One can see that, in order to get the correct DM relic for a 1 GeV DM, the required decay width $\Gamma_{N_R \to \phi \chi} = 3\times 10^{-20}$ GeV turns out to be smaller (bigger) than the ones in case-I (case-II) depicted in the left (right) panel of Fig.~\ref{relic1}. 

Thus, Fig. \ref{relic_scan} explains the role of freeze-out cross section $\langle\sigma v \rangle^{^{N_R}}_{_{ \rm F.O.}}$ and decay width $\Gamma_{N_R \to \phi \chi}$ of $N_R$ in generating the relic abundance of DM. In order to see the dependence on $N_R$ mass, we show two different benchmark values of $N_R$ mass namely $M_{N_R} =100$ GeV and $M_{N_R} =1000$ GeV. When $N_R$ mass decreases by a factor of 10, the required freeze-out cross section, increases by same factor to generate required relic of DM having mass 1 GeV, as seen from comparison of points marked as $B$ ($m_\chi = 1 \,{\rm GeV},
M_{N_R} = 1000 \,{\rm GeV}$ and $\langle\sigma v \rangle^{^{N_R}}_{_{ \rm F.O.}}= {\cal O}(10^{-11}/{\rm GeV}^2$) 
and $B'$ ($m_\chi = 1 \,{\rm GeV}, M_{N_R} = 100 \,{\rm GeV}$ and $\langle\sigma v \rangle^{^{N_R}}_{_{ \rm F.O.}}= {\cal O}(10^{-10}/{\rm GeV}^2$) on left panel of Fig. \ref{relic_scan}. Such small decay width regime requires freeze-out abundance of $N_R$ to be finely tuned to the required DM abundance, as seen from the right panel plot of Fig. \ref{relic1}. Since freeze-out number density of $N_R$ is inversely proportional to its mass as well as freeze-out cross section, one can understand the increase in freeze-out cross section while moving from point $B$ to $B'$ on the left panel plot of Fig. \ref{relic_scan}. On the other hand, if we focus on the region with low freeze-out cross section, the contours for $M_{N_R}=1000$ GeV and $M_{N_R}=100$ GeV merge with each other. This is simply because, for such low freeze-out cross-section, $N_R$ abundance will be much more than DM abundance and hence the required DM relic can be generated only by ensuring subsequent DM annihilation into $Z'$ pairs. This also reduces the dependence of final DM relic on freeze-out cross section as evident from the horizontal portion of the contours on left panel plot of Fig. \ref{relic_scan}. This feature is also shown on left panel plot of Fig. \ref{relic1} in terms of the red solid line representing the evolution of comoving DM number density. On the left panel plot of Fig. \ref{relic_scan}, the point marked as $A (A')$ on the cyan (green) coloured line corresponds to that value of freeze-out cross-section below which the correct DM relic can be generated by the same decay width at that point. However, the freeze-out cross section can not be arbitrarily lower, as $N_R$ needs to be thermally produced in the universe. Similar behaviour is noticed for a larger value of DM mass $m_{\chi}=10$ GeV as shown by blue and red dashed lines on the left panel plot of Fig. \ref{relic_scan}. Compared to lighter DM, here the contours shift towards up and right, if $N_R$ mass is kept fixed. This is primarily because for heavier DM, the required number density is less and hence a larger freeze-out cross section of $N_R$ is also consistent in the rightmost part of the plot. For small freeze-out cross section, the corresponding decay width also becomes more (compared to the ones for lighter DM) because, heavy DM annihilation into $Z'$ pairs will become inefficient at earlier epochs and hence it is important to ensure that $N_R$ decays into DM at much earlier epoch. The points marked as $C, C'$ can be understood in an analogous manner to points marked as $B, B'$ discussed above. On the right panel plot of Fig. \ref{relic_scan}, we show the allowed region in $M_{N_R}-\langle\sigma v \rangle^{^{N_R}}_{_{ \rm F.O.}}$ plane by considering fixed $N_R-\chi$ coupling $y=10^{-10}$ and DM mass $m_{\chi}=1$ GeV. The behaviour can be understood in a way similar to the left panel plot, by noting that the decay width of $N_R$ is proportional to its mass.

Now, let us discuss Fig.~\ref{relic1}, \ref{relic2} mentioned earlier where we show the evolution of comoving number densities of different species for three cases. The magenta dashed lines in all these figures show the freeze-in production of DM from the SM bath (electrons, for example) without considering any other processes like subsequent DM annihilation, RHN decay. This occurs due to the kinetic mixing of $Z'$ with SM gauge bosons and we choose the corresponding parameter to be $\epsilon = 10^{-9}$. While DM abundance from freeze-in is huge compared to observed relic, incorporating subsequent DM annihilation into $Z'$ pairs leads to much smaller DM relic, shown by the blue dashed lines. Due to the competing processes of DM production from freeze-in and subsequent DM annihilations, the blue dashed lines show rise, fall as well as plateau regions followed by a dark sector freeze-out around $x \sim 13$ leaving a saturated but under-abundant DM relic shown by the blue dashed line at large values of x\footnote{It is to be noted that since DM annihilation into light mediator is very large, final abundance after dark sector freeze-out does not significantly depends on initial freeze-in abundance from processes like $e^{+}e^{-} \to \chi \chi$ (which is proportional to $\epsilon^2$).}. On the other hand, $N_{R}$ can be produced in equilibrium from the SM bath at very early epochs with the brown dashed line showing its equilibrium abundance, while the orange dashed line indicates its freeze-out abundance, leaving a sizeable relic depending on $\langle\sigma v \rangle^{^{N_R}}_{_{ \rm F.O.}}$. At late epochs $N_R$ decays into DM filling the deficit in DM relic with the corresponding depletion in $N_R$ abundance is shown by the solid green line. The corresponding increase in DM abundance is shown by the solid red line in Fig.~\ref{relic1}, \ref{relic2}.


\subsection{Direct Detection}
\label{dd}
\begin{figure}[h!]
\centering
\includegraphics[scale=0.20]{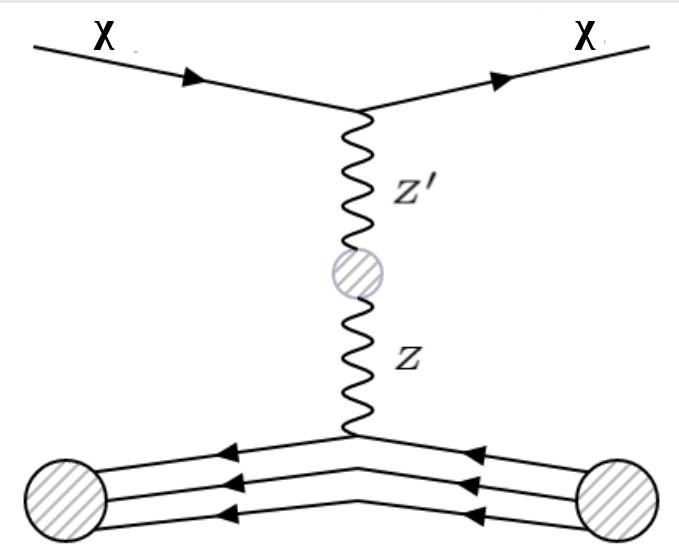}
\caption{Elastic scattering of DM $\chi$ off a nucleon mediated by $Z-Z'$ mixing.}
\label{direct}
\end{figure}

\begin{figure}[h!]
\centering
\includegraphics[scale=0.45]{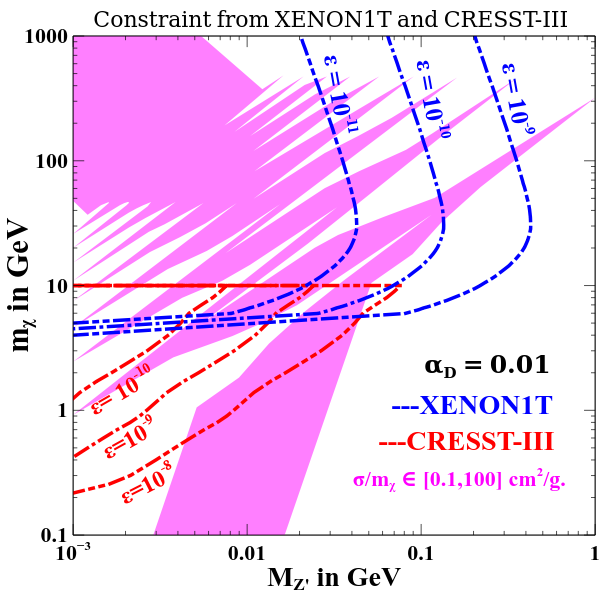}
\hfil
\includegraphics[scale=0.45]{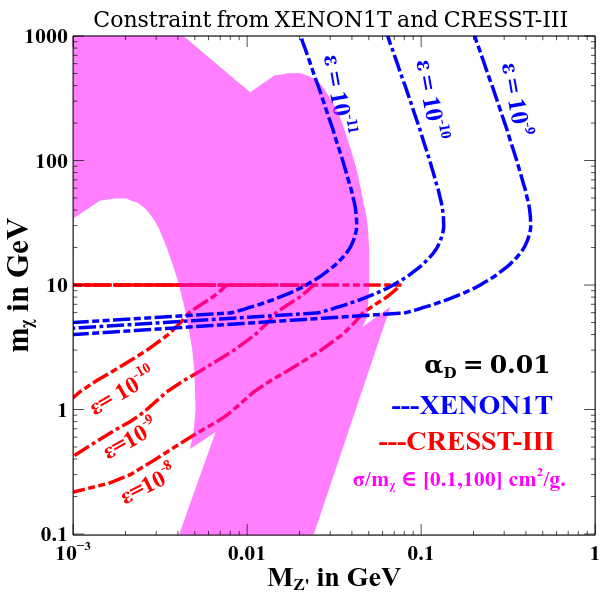}
\caption{Constraints from DM direct detection in the plane of DM mass $(m_{\chi})$ versus mediator mass $(M_{Z'})$ for attractive (left panel) and repulsive (right panel) self-interactions respectively.}
\label{sidmdd}
\end{figure}

Direct detection of dark matter $\chi$ is possible through elastic scattering of $\chi$ with the nucleons of the detector atoms through $Z-Z'$ kinetic mixing\footnote{Since kinetic mixing is between $U(1)_D$ and $U(1)_Y$, $Z'$ can mix with photon as well but we consider Z-Z' only due to larger Z coupling of SM.} as shown in Fig.~\ref{direct}. The Feynman diagram for the spin independent direct search cross section for a given nucleus N with proton number Z and mass number A is given by
\begin{equation}
\sigma^{\rm SI}_{\chi N}=\frac{g^2 g^2_D \epsilon^2}{\pi}\frac{ \mu^2_{\chi N}}{M^4_{Z'}} \frac{\big( Z f_p + (A-Z) f_n \big)^2}{A^2}.
\end{equation}
Here, $\mu_{\chi N} = \frac{m_\chi m_N}{(m_\chi+m_N)}$ is the reduced mass of the DM-nucleus system, $\epsilon$ is the $Z-Z'$ kinetic mixing, $f_p$ and $f_n$ are the are the interaction strengths for proton and neutron respectively.
As mentioned earlier, DM direct search experiments like CRESST-III~\cite{Abdelhameed:2019hmk} and XENON1T \cite{Aprile:2018dbl} can constrain the model parameters. While XENON1T provides the most stringent bound for DM of mass above 10 GeV, CRESST constraints the below 10 GeV mass range. In Fig.~\ref{sidmdd}, the most stringent constraints from CRESST-III~\cite{Abdelhameed:2019hmk}, XENON1T \cite{Aprile:2018dbl} experiments on $m_\chi-m_{Z'}$ plane are shown against the parameter space favoured from required DM self-interactions by assuming $\alpha_D=0.01$. The left panel plot of Fig.~\ref{sidmdd} shows the case of attractive self-interaction while the right panel shows repulsive one. The blue coloured contours denote exclusion limits from XENON1T experiment for specific kinetic mixing parameters such that the region towards the left of the contour is excluded. Similarly, the red coloured contours show the CRESST-III bound on low mass DM for different kinetic mixing parameters ruling out the parameter space towards left of each contour.

\subsection{Summary: Minimal Setup}
\begin{figure}[h!]
	\centering
	\includegraphics[scale=0.45]{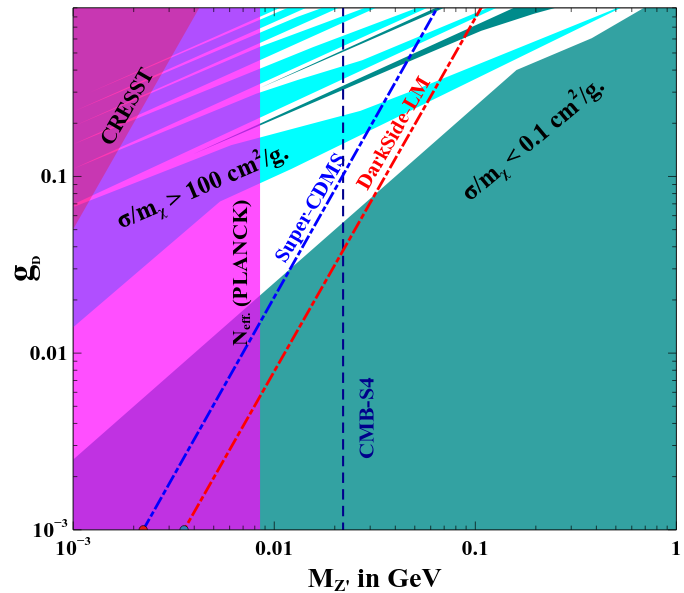}
	\hfil
	\includegraphics[scale=0.45]{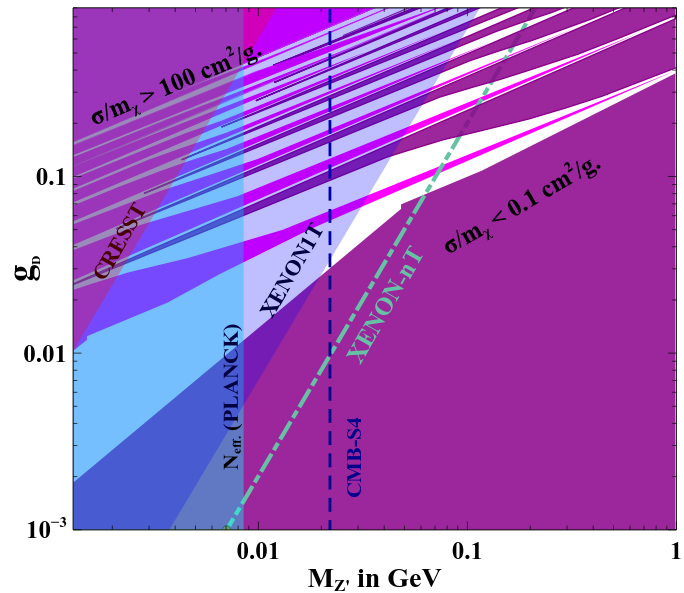}
	\caption{Summary plot showing parameter space in $g_D-M_{Z'}$ plane for $m_{\chi}=1$ GeV (left) and $m_{\chi}=10$ GeV (right).}
	\label{summary10}
\end{figure}

\begin{figure}[h!]
	\centering
	\includegraphics[scale=0.55]{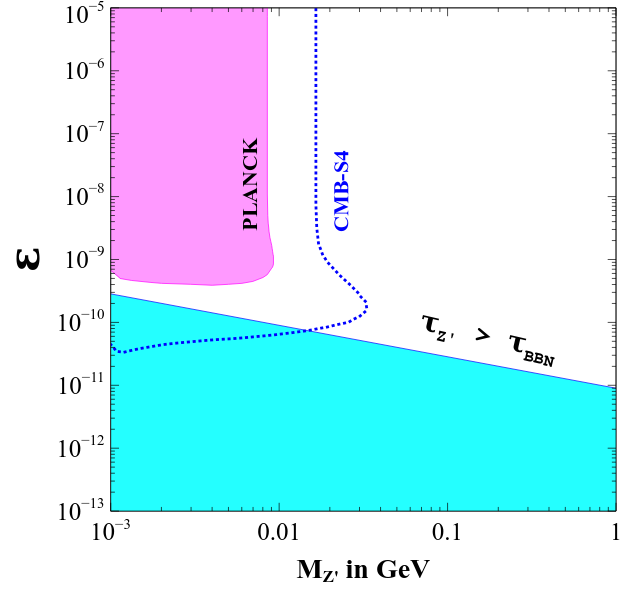}
	\caption{Kinetic mixing parameter $\epsilon$ versus $Z'$ mass showing the region of parameter space where $Z'$ can be long-lived and decay after BBN. The magenta region shows the region of parameter space excluded from cosmological constraints on effective relativistic degrees of freedom. The Blue line shows the projected sensitivity of CMB-S4 experiment\cite{CMB-S4:2016ple,Ibe:2019gpv}.}
	\label{summary10a}
\end{figure}

In Fig. \ref{summary10}, we summarize the parameter space in $g_D-M_{Z'}$ plane for two benchmark DM masses $m_{\chi}=1$ GeV, $m_{\chi}=10$ GeV on left and right panels respectively. The upper left and lower right regions are disfavoured as they give rise to too large and too small DM self-interactions respectively, leaving a band in between.  Very light $Z'$ below a few MeV is ruled out from cosmological constraints on effective relativistic degrees of freedom \cite{Aghanim:2018eyx, Kamada:2018zxi, Ibe:2019gpv, Escudero:2019gzq}. This constraint has been shown by the Magenta shaded region in Fig.~\ref{summary10a}. This arises due to the late decay of such light gauge bosons into SM leptons, after standard neutrino decoupling temperatures thereby enhancing $N_{\rm eff}$. We also showcase the parameter space sensitive to the future sensitivity of CMB-S4 experiment\cite{CMB-S4:2016ple,Ibe:2019gpv}. This constraint has also been shown in Fig.~\ref{summary10} by the dark blue dotted line. In Fig.~\ref{summary10}, a small portion of the upper left triangular region in left panel plot is disfavoured by CRESST-III bound on low mass DM while a large portion of the right panel plot is ruled out by XENON1T bound for fixed kinetic mixing $\epsilon = 10^{-10}$. The blue and red dotted lines in the left panel depicts the projected sensitivity of SuperCDMS~\cite{SuperCDMS:2016wui} and DarkSide-LM~\cite{DarkSide:2018bpj} for DM mass 1 GeV and similarly the green dotted line in the right panel plot shows the projected sensitivity of XENON-nT~\cite{XENON:2020kmp} experiment for DM mass 10 GeV~\cite{EuropeanStrategyforParticlePhysicsPreparatoryGroup:2019qin}. Since DM relic can be satisfied independently by fixing parameters related to RHN, it does not impose any further constraints in the plots shown in Fig. \ref{summary10}. Since late DM annihilations produce $Z'$ pairs copiously, we apply a conservative bound on $Z'$ lifetime to be less than typical BBN epoch so as not to disturb the predictions of light nuclei abundance by injecting entropy. In Fig. \ref{summary10a}, kinetic mixing parameter $\epsilon$ is shown against $M_{Z'}$. Very small kinetic mixing (the cyan coloured region) is disfavoured from this conservative lifetime bound. The chosen benchmark values of $\epsilon$ in above discussion and results automatically satisfy this lower bound. Note that due to large gauge coupling, decay of $\phi \to Z'Z'$ happens well before BBN. The decay width of these processes are summarized in Appendix~\ref{appen1}.

It should be noted that we are considering the kinetic mixing parameter $\epsilon$ to be a free parameter in the above discussion. However, it can be generated at one-loop level as well. There are only two particles namely $\chi, \Phi$ which are charged under $U(1)_D$ and out of them, the DM $\chi$ does not have any direct coupling to SM particles. Therefore, the radiative generation of kinetic mixing is possible via scalar loop, suppressed by singlet scalar mixing with the SM Higgs $\theta_{\phi h}$ as \cite{Cheung:2009qd, Mambrini:2011dw}
\begin{equation}
    \epsilon_{\rm 1\, loop} \propto \frac{g_D g_Y}{16\pi^2} \theta^2_{\phi h},
\end{equation}
where $g_Y$S is the corresponding $U(1)_Y$ gauge coupling. Since the rest of the phenomenology does not depend upon singlet-SM Higgs mixing, we can tune it arbitrarily to keep the one-loop kinetic mixing suppressed compared to the leading order value considered in the numerical analysis.

In addition to the DM phenomenology discussed above, the singlet scalar $\Phi$ can also have interesting phenomenology. While it does not couple to the SM fields directly, it can do so via its mixing with the SM Higgs. For light $Z'$ masses and order one gauge coupling $g_D$, the singlet scalar VEV is expected to be of similar order as the $M_{Z'}$ resulting in light mass of the physical singlet scalar. If the singlet scalar has sizeable mixing with the SM Higgs, it can be produced in thermal bath. Since the Yukawa coupling for $N_R-\chi$ interaction with $\Phi$ is considered to be tiny for late decay of $N_R$, one can also have freeze-in production of DM via scalar portal interactions. In the above discussion, we have considered only gauge portal channels to be the dominant production channel for DM for simplicity and adding scalar portal interactions should not change the phenomenology significantly. The light singlet scalar can of course have other interesting phenomenology which can be probed at experiments \cite{Clarke:2013aya}.

As discussed above, the Yukawa coupling of DM with heavier RHNs $(N_{2,3})$ are assumed to be absent for simplicity although the symmetries of the model can not forbid them. Considering them to be non-zero but as small as $N_{1}$ coupling will not make any difference to the results as $N_{2, 3}$ are required to have larger Yukawa couplings with leptons from neutrino mass constraints and hence they will decay into SM particles preferentially. Considering large Yukawa couplings of $N_{2,3}$ with DM can, however, lead to thermal production of DM. But since DM continues to annihilate strongly into $Z'$ pairs, the late decay of $N_1$ is still required to fill up the thermal deficit, keeping the overall scenario similar to what we have discussed here.

\section{UV Complete Realization: Two Scenarios}
\label{examples}
After discussing the phenomenology of right handed neutrino portal self-interacting DM, we now outline two specific UV complete realizations which can not only give rise to required RHN thermal abundance but also connects to the origin of light neutrino masses. We impose the constraints on RHN abundance from self-interacting DM relic criteria and find the specific model parameter space. We also compare them with existing experimental constraints and check their verifiability in near future experiments at different frontiers.
\subsection{Scotogenic Realization} 
We first consider the implementation of RHN portal SIDM scenario along with the scotogenic model. As proposed by Ma in 2006 \cite{Ma:2006km}, scotogenic model is an extension of the SM by three copies of RHNs and an additional scalar doublet $\eta$ all of which are odd under an in-built $Z_2$ symmetry. While the SM fields remain even under $Z_2$, it is still possible for SM lepton doublets to couple to $\eta$ and RHNs giving rise to the possibility of radiative neutrino masses. The BSM particle content of the model are shown in table \ref{tab:tab1}. Apart from the usual $Z_2$ and $U(1)_D$ sectors, note that the fermion DM $\chi$ is also odd under $Z_2$ symmetry in order to allow its RHN portal coupling with singlet scalar $\Phi$.

\begin{table}[h!]
	\begin{tabular}{|c|c|c|c|}
		\hline \multicolumn{2}{|c}{Fields}&  \multicolumn{1}{|c|}{ $SU(3)_c \otimes SU(2)_L \otimes U(1)_Y$ $\otimes~~  U(1)_D \otimes~~Z_2$  } \\ \hline
		{Fermion} &  $N_R$&  ~~1 ~~~~~~~~~~~1~~~~~~~~~~0~~~~~~~~~~~0~~~~~~~~~ - \\ 
		& $\chi$  & ~~1 ~~~~~~~~~~~1~~~~~~~~~~0~~~~~~~~~~~1~~~~~~~~~ - \\
		[0.5em] \cline{2-3}
		\hline
		Scalars & $\eta$ &~~1 ~~~~~~~~~~~2~~~~~~~~~~1~~~~~~~~~~~0~~~~~~~~~ - \\
		& $\Phi$ & ~~1 ~~~~~~~~~~~1~~~~~~~~~~0~~~~~~~~~~~-1~~~~~~~~~ + \\
		\hline
	\end{tabular}
\caption{BSM particle content and their transformation under the chosen symmetry in scotogenic realization of RHN portal SIDM.}
\label{tab:tab1}
\end{table}
The relevant Lagrangian for neutrino mass generation consistent with the imposed symmetry can be written as
\begin{equation}
\mathcal{L} \supset -\frac{1}{2}M_{N_R} \overline{N^c_R} N_R - Y_N \overline{L} \tilde{\eta} N_R +  {\rm h.c.}
\end{equation}
The scalar potential involving the new scalar doublet $\eta$ is.
\begin{eqnarray}
V(H,\eta)&=&-\mu^2_H H^\dagger H + \frac{\lambda_H}{2}(H^\dagger H)^2+\mu^2_\eta \eta^\dagger \eta + \frac{\lambda_\eta}{2}(\eta^\dagger \eta)^2\nonumber\\&+&\lambda_{H \eta} (H^\dagger H)(\eta^\dagger \eta)+\lambda'_{H \eta} (H^\dagger \eta)(\eta^\dagger H)+\frac{\lambda^{''}_{H \eta}}{2}\big[(H^\dagger \eta)^2+(\eta^\dagger H)^2\big]
\end{eqnarray}
\begin{figure}[h!]
\centering
\includegraphics[scale=0.45]{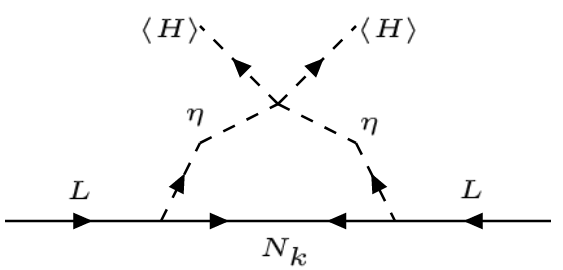}
\caption{Generation of one-loop neutrino mass in scotogenic model.}
\label{scoto}
\end{figure}
After the electroweak symmetry breaking (EWSB), these two scalar doublets can be written in unitary gauge as
\begin{equation}
H \ = \ \begin{pmatrix} 0 \\  \frac{ v +h }{\sqrt 2} \end{pmatrix} , \qquad \eta \ = \ \begin{pmatrix} H^\pm\\  \frac{H^0+iA^0}{\sqrt 2} \end{pmatrix} \, ,
\label{eq:idm}
\end{equation}
where $v$ is the VEV of SM Higgs $H$ responsible for EWSB.
The neutral scalar and pseudoscalars $H^0, A^0$ acquire masses as follows.
\begin{equation}
M^2_{R,I}=\mu^2_\eta+\frac{1}{2}(\lambda_{H \eta}+\lambda'_{H \eta}\pm \lambda''_{H \eta})~v^2,
\end{equation}
with $+ (-)$ denoting scalar (pseudoscalar). And the neutrino mass induced via the one-loop diagram as shown in Fig.~\ref{scoto} is given by
\begin{eqnarray}
(\mathcal{M}_\nu)_{\alpha \beta}=\sum_{k=1}^{3} \frac{(Y_{N})_{k \beta} (Y_{N})_{k \alpha}}{32 \pi^2} M_{Nk}\bigg[\frac{M^2_R}{M^2_R-M^2_{Nk}}\ln \bigg(\frac{M^2_R}{M^2_{Nk}}\bigg)-\frac{M^2_I}{M^2_I-M^2_{Nk}}\ln \bigg(\frac{M^2_I}{M^2_{Nk}}\bigg)\bigg]
\end{eqnarray}

where $M_{N_k}$ is the mass eigenvalue of the RHN mass eigen-
state $N_k$ in the internal line and the indices $\alpha, \beta = e, \mu, \tau$
run over the three neutrino generations.
Neutrino
mass vanishes in the limit of $\lambda''_{H \eta} \to 0$ as it corresponds to degenerate neutral scalar and pseudoscalar masses $M^2_R=M^2_I$. Thus, apart form the Yukawa coupling $Y_N$ and RHN masses, this quartic coupling also play significant role in neutrino mass generation. To include the constraints from light neutrino data in the analysis, it is often convenient to write the Yukawa couplings in Casas-Ibarra parametrisation \cite{Casas:2001sr,Toma:2013zsa} as
\begin{equation}
Y = {\sqrt{\Lambda}}^{-1} R \sqrt{\hat{m_\nu}} U^\dagger_{\rm PMNS} 
\end{equation}
where $R$ is an arbitrary complex orthogonal matrix satisfying $RR^{T}=\mathrm{I}$ that can be parameterized in terms of three complex angles ($\alpha$, $\beta$, $\gamma$). Here, $\hat{m_\nu} =  \textrm{Diag}(m_1,m_2,m_3)$ is the diagonal light neutrino mass matrix and the diagonal matrix $\Lambda$ is defined as $\Lambda$ = Diag ($\Lambda_1$,$\Lambda_2$,$\Lambda_3$), with
\begin{equation}
\Lambda_k=\frac{M_{N_k}}{32 \pi^2} \bigg[\frac{M^2_R}{M^2_R-M^2_{N_k}}ln \bigg(\frac{M^2_R}{M^2_{N_k}}\bigg)-\frac{M^2_I}{M^2_I-M^2_{N_k}}ln \bigg(\frac{M^2_I}{M^2_{N_k}}\bigg)\bigg].
\end{equation}
$U_{\rm PMNS}$ is the usual Pontecorvo-Maki-Nakagawa-Sakata (PMNS) mixing matrix of neutrinos. 

\subsubsection{Lightest RHN Abundance}
To realize the RHN portal SIDM scenario within the scotogenic framework, we consider the lightest RHN $(N_1)$ to be lighter than $\eta$ so that it can not decay into the SM particles. And since $N_1$ coupling with SIDM is fine-tuned, it can be long-lived. Thus, although $N_1$ is not DM in our model, it can still undergo thermal freeze-out leaving a relic which later gets converted into actual DM, in a way similar to super-WIMP DM scenario \cite{Feng:2003uy}. As discussed in subsection \ref{dm_production}, for a particular benchmark of SIDM parameter space, one requires a certain freeze-out abundance of $N_1$ so that correct SIDM relic abundance is generated at the end from non-thermal decay of $N_1$. Therefore, we estimate the thermal freeze-out abundance of $N_1$ in the scotogenic model in a way similar to fermion DM studies in scotogenic model \cite{Ahriche:2017iar, Mahanta:2019gfe, Borah:2020wut}.

Since $N_1$ interacts with the SM bath only via Yukawa couplings, one requires sizeable Yukawas in order to generate the required thermal relic. Unlike in conventional seesaw models at low scale, scotogenic model offers the possibility to have sizeable Yukawa couplings and still satisfy the light neutrino mass because of the loop suppression involved and the freedom to choose the scalar quartic coupling. Depending upon the size of Yukawa couplings as well as mass splitting with other $Z_2$ odd particles, required $N_1$ abundance can be generated due to dominance of usual annihilation, coannihilation \cite{Griest:1990kh} or a combination of both. And these Yukawa couplings are essentially instrumental in keeping the RHNs in thermal equilibrium in early universe which then decouples and decays to DM $\chi$ and $\phi$ to generate the correct DM relic through a non-thermal mechanism as discussed in section~\ref{dm_production}. While singlet RHNs are not constrained from collider searches, the requirement of large Yukawa couplings can lead to some tensions with other experimental bounds like charged lepton flavour violation which we discuss below.

\subsubsection{Lepton flavour violation}
\label{lfv}
In the SM, charged lepton flavour violating (CLFV) decays occurs at loop level and is highly suppressed by the smallness of neutrino masses, and thus is much beyond the current experimental sensitivity \cite{TheMEG:2016wtm}. Therefore, any future observation of such LFV decays like $\mu \rightarrow e \gamma$ will definitely be an indication of new physics beyond the SM. 

\begin{figure}[h!]
\centering
\includegraphics[scale=0.55]{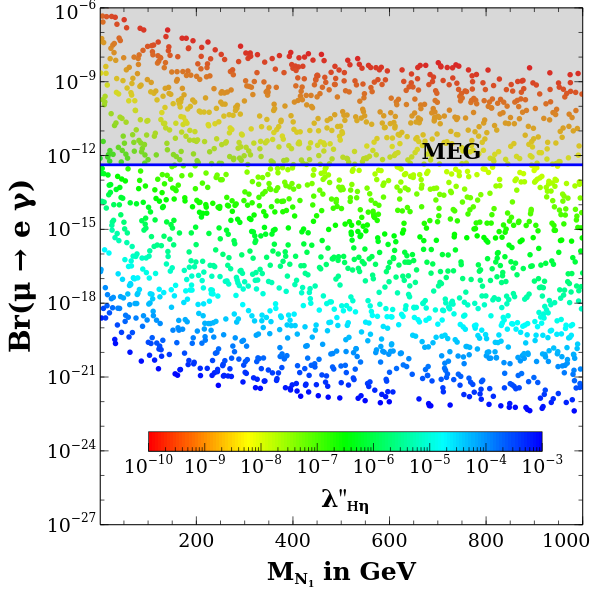}
\caption{Br($\mu \rightarrow e \gamma$) as a function of $N_1$ mass while keeping $M_{N_1}=M_{N_2}-10 \, {\rm GeV} = M_{N_3}-20 \, {\rm GeV}, M_{\eta^+} = 1$ TeV.}
\label{scotolfv}
\end{figure}

In the scotogenic model, the charged component of the additional scalar doublet $\eta$ going inside a loop along with singlet fermions can facilitate such CLFV decays. 

{\underline{ $\pmb{\mu\to e \gamma}$}}

Following the prescriptions given in \cite{Lavoura:2003xp, Toma:2013zsa}, the decay width of $\mu \rightarrow e \gamma$ can be calculated as
\begin{align}
{\rm Br} (\mu \rightarrow e \gamma) =\frac{3 (4\pi)^3 \alpha}{4G^2_F} \lvert A_D \rvert^2 {\rm Br} (\mu \rightarrow e \nu_{\mu} \overline{\nu_e}),
\end{align}
where $A_D$ is given by
\begin{equation}
A_D = \sum_{k} \frac{(Y_N)^*_{ke} (Y_N)_{k\mu}}{16 \pi^2} \frac{1}{M^2_{\eta^+}} f (r_k),
\label{ADMEG}
\end{equation}
with $r_k = M^2_{N_k}/M^2_{\eta^+}$. $f(x)$ is the loop function given by
\begin{equation}
f(x)=\frac{1-6x+3x^2+2x^3-6x^2\log{x}}{12 (1-x)^4}.
\label{loop1}
\end{equation}

The latest bound from the MEG collaboration is $\text{Br}(\mu \rightarrow e \gamma) < 4.2 \times 10^{-13}$ at $90\%$ confidence level \cite{TheMEG:2016wtm}. 
In Fig.~\ref{scotolfv}, Br($\mu \rightarrow e \gamma$) is shown as a function of $N_1$ mass. For the scan, we fixed the mass splitting between $N_1$ with $N_2$ and $N_3$ at $10$ GeV and $20$ GeV respectively and fixed charged scalar $\eta^+$ mass at $1$ TeV.  Here considering the normal ordering, the lightest active neutrino mass is assumed to be $10^{-3}$eV consistent with the constraint for the sum of active neutrino masses $\sum m_{\nu}= 0.12$ eV from cosmological data. The solid Blue line represents the latest upper limit from the MEG experiment. It is clear that $\lambda''_{H\eta}$ smaller than $\sim \mathcal{O}(10^{-8})$ is disfavoured from the CLFV constraint consequently putting a upper limit on the Yukawa couplings.

\underline{ $\pmb{\mu\to 3 ~e}$}

	In addition to $\mu\to e\gamma$, another CLFV observable that can provide distinctive signature is the three body decay process $\mu \to 3e $. This branching fraction is given by~\cite{Toma:2013zsa}:
\begin{eqnarray}
	\text{Br}\left(\mu \to
	e \overline{e}e\right)&=&
	\frac{3(4\pi)^2\alpha_{\mathrm{em}}^2}{8G_F^2}
	\left[|A_{ND}|^2
	+|A_D|^2\left(\frac{16}{3}\log\left(\frac{m_\mu}{m_e}\right)
	-\frac{22}{3}\right)+\frac{1}{6}|B|^2\right.\nonumber\\
	&&\left.+ \frac{1}{3} \left( 2  |F_{RR}|^2 + |F_{RL}|^2 \right)
	+\left(-2 A_{ND} A_D^{*}+\frac{1}{3}A_{ND} B^*
	-\frac{2}{3}A_D B^*+\mathrm{h.c.}\right)\right]\nonumber\\
	&&\times \, \mathrm{Br}\left(\mu \to e\nu_{\mu}
	\overline{\nu_e}\right) \,  \label{eq:l3lBR}
\end{eqnarray}
where $A_D$ is as given in Eq.~\ref{ADMEG}, and $A_{ND}$ is given by:
\begin{equation}
	A_{ND}=\sum_{k=1}^3\frac{(Y_N)^*_{ke} (Y_N)_{k\mu}}
	{6(4\pi)^2}\frac{1}{M_{\eta^+}^2}
	G_2\left(r_k\right), \label{eq:A1L}
\end{equation}
and 
\begin{equation}
	F_{RR} = \frac{F \, g_R^\ell}{g_2^2 \sin^2 \theta_W M_Z^2} \qquad ,
	\qquad F_{RL} = \frac{F \, g_L^\ell}{g_2^2 \sin^2 \theta_W M_Z^2} \quad,
\end{equation}
with the co-efficient $F$ given by
\begin{equation}
	F = \sum_{k=1}^3\frac{(Y_N)^*_{ke} (Y_N)_{k\mu}}
	{2(4\pi)^2}\frac{m_\mu m_e}{M_{\eta^+}^2} \frac{g_2}{\cos \theta_W}
	F_2\left(r_k\right) \, .
	\label{eq:FR}
\end{equation}
For the Box diagrams, the co-efficient $B$ is given by:
{\small \begin{align}
	 B = \frac{1}{(4\pi)^2 e^2 M_{\eta^+}^2} 
	\sum_{j,\:k=1}^3\left[\frac{1}{2} D_1(r_j,r_k) (Y_N)_{k e}^* (Y_N)_{k e}
	(Y_N)_{j e}^* (Y_N)_{j \mu} + \sqrt{r_j r_k}
	D_2(r_j,r_k) (Y_N)_{k e}^* (Y_N)_{k e}^* (Y_N)_{je}
	(Y_N)_{j \mu}\right].
\end{align}}
The loop functions $G_2, F_2, D_1, D_2$ are given in Appendix \ref{loopfunc}.
\begin{figure}[h!]
	\centering
	\includegraphics[scale=0.5]{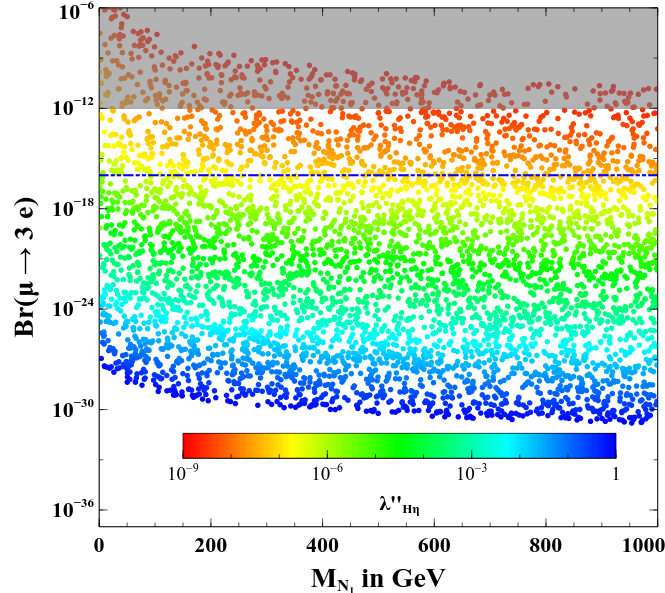}
	\caption{Br($\mu \to 3 e$) as a function of $N_1$ mass while keeping $M_{N_1}=M_{N_2}-10 \, {\rm GeV} = M_{N_3}-20 \, {\rm GeV}, M_{\eta^+} = 1$ TeV. The grey shaded region shows the present bound~\cite{SINDRUM:1987nra} and the blue dotted line shows the future sensitivity.~\cite{Baldini:2018uhj}}
	\label{muto3e}
\end{figure}

\underline{ $\pmb{\mu\to e}$ \bf{conversion in nuclei}}

	Because of the great projected sensitivities of various collaborations,  
	$\mu \to e$ conversion in nuclei might become the most severely constrained observable in scotogenic scenarios.
The conversion rate, relative to the
the muon capture rate, can be expressed as:
\begin{align}
	{\rm CR} (\mu- e, {\rm Nucleus}) &= 
	\frac{p_e \, E_e \, m_\mu^3 \, G_F^2 \, \alpha_{\mathrm{em}}^3 
		\, Z_{\rm eff}^4 \, F_p^2}{8 \, \pi^2 \, Z}  \nonumber \\
	&\times \left\{ \left| (Z + N) \left( g_{LV}^{(0)} + g_{LS}^{(0)} \right) + 
	(Z - N) \left( g_{LV}^{(1)} + g_{LS}^{(1)} \right) \right|^2 + 
	\right. \nonumber \\
	& \ \ \ 
	\ \left. \,\, \left| (Z + N) \left( g_{RV}^{(0)} + g_{RS}^{(0)} \right) + 
	(Z - N) \left( g_{RV}^{(1)} + g_{RS}^{(1)} \right) \right|^2 \right\} 
	\frac{1}{\Gamma_{\rm capt}}\,.
\end{align}
where $Z$ and $N$ are the number of protons and neutrons in the nucleus, $Z_{\rm eff}$ is the effective atomic
charge, $F_p$ is the nuclear matrix element and $\Gamma_{\rm capt}$ represents the total muon capture rate. Furthermore, $p_e$ and $E_e$ (taken to be $\simeq m_\mu$ in the numerical evaluation) are
the momentum and energy of the electron and $m_\mu$ is the muon mass. The other couplings can be found in Appendix~\ref{appen3}.
\begin{figure}[h!]
	\centering
	\includegraphics[scale=0.5]{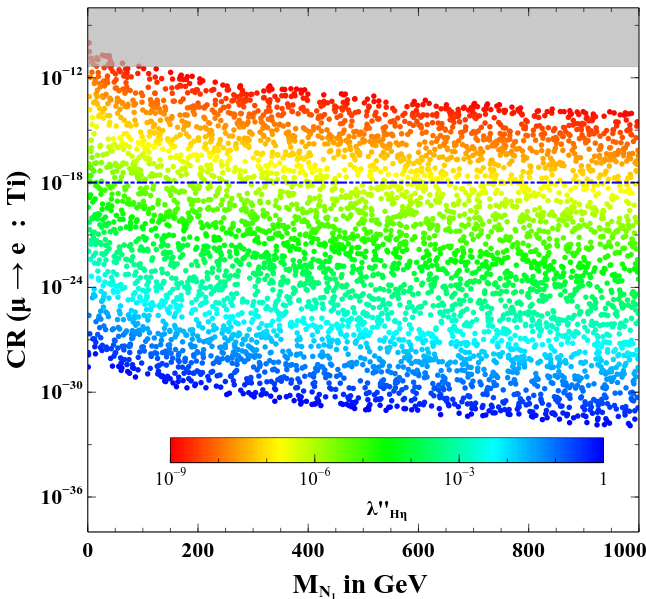}
	\caption{CR($\mu \to e : {\rm Ti}$) as a function of $N_1$ mass while keeping $M_{N_1}=M_{N_2}-10 \, {\rm GeV} = M_{N_3}-20 \, {\rm GeV}, M_{\eta^+} = 1$ TeV. The grey shaded region shows the present bound~\cite{SINDRUMII:1993gxf} and the blue dotted line shows the future sensitivity.~\cite{Baldini:2018uhj}}
	\label{mutoeconv}
\end{figure}

\begin{figure}[h!]
\centering
\includegraphics[height=7cm,width=7cm]{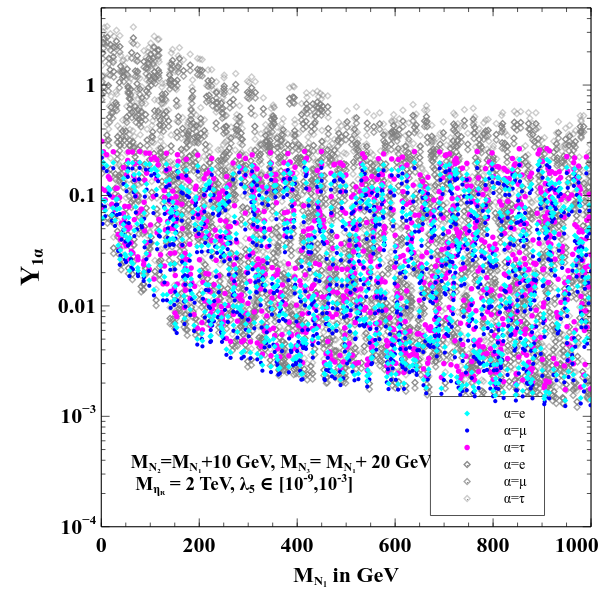}
\hfil
\includegraphics[height=7cm,width=7cm]{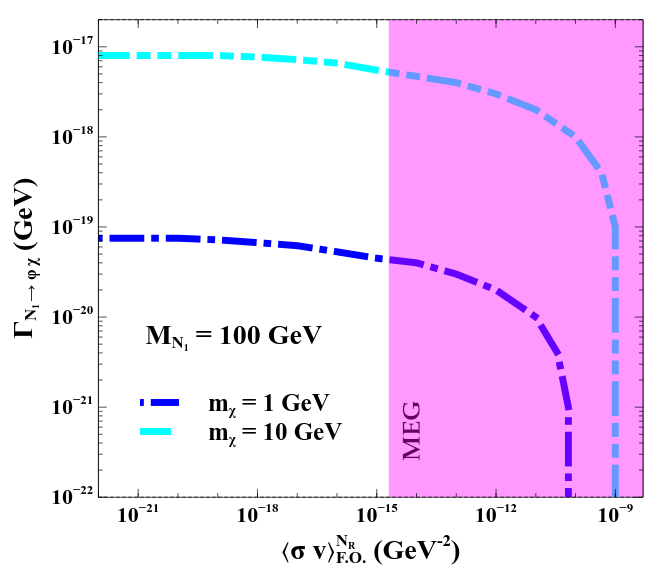}
\caption{[Left]:Parameter space for scotogenic model in the plane of $N_1$ Yukawa and its mass consistent with SIDM relic. The coloured dots (blue, cyan, pink) are allowed from CLFV bounds while the grey coloured box shaped points which do not overlap with the colored points are disallowed.
	[Right]: Contours of correct DM relic in the plane of $\langle\sigma v \rangle^{^{N_1}}_{_{ \rm F.O.}}$ and $\Gamma_{N_1 \to \phi \chi}$ where the magenta shaded region is disfavoured due to CLFV bounds from the MEG experiment. }
\label{scotops}
\end{figure}

Having studied all these CLFV processes, finally we show the scotogenic model parameter space consistent with SIDM relic criteria in left panel of Fig. \ref{scotops}  in the plane of $N_1$ Yukawa and its mass. As mentioned earlier, for a fixed range or value of SIDM parameters, we have a constraint on the cross-section $\langle\sigma v \rangle^{^{N_1}}_{_{ \rm F.O.}}$ such that the required $N_1$ freeze-out abundance is generated which later gets transferred into SIDM $\chi$.
In this figure, the points in blue, cyan and pink coloured dots are allowed from CLFV bounds while the grey coloured box shaped points which do not overlap with the coloured points are disallowed. This clearly shows that the model parameters consistent with SIDM relic remains verifiable at ongoing experiments. In the right panel of Fig.~\ref{scotops}, the contours of correct DM relic in the plane of $\langle\sigma v \rangle^{^{N_1}}_{_{ \rm F.O.}}$ and $\Gamma_{N_1 \to \phi \chi}$ are shown for $M_{N_1}=100$ GeV and $m_{\chi}=1,10$ GeV similar to the left panel plot of Fig.~\ref{relic_scan}. Here the magenta shaded region depicts the region disfavoured by CLFV bounds from the MEG experiment \cite{TheMEG:2016wtm} which is the most stringent among all CLFV constraints. Since the freeze-out cross-section of $N_1$ depends upon its Yukawa couplings with leptons, the disfavoured region corresponds to large cross section and hence large Yukawa couplings which give a very large CLFV rate disfavoured by experimental bounds. For this analysis, we have ignored the co-annihilation effects by maintaining a minimum mass difference of $100$ GeV between lightest RHN $(N_1)$ and the doublet scalar $\eta$. However we have checked that if this mass-difference is below $30$ GeV, then it can enhance the effective freeze-out cross-section of $N_1$ because of co-annihilations relaxing the dependence as well as bound on the Yukawa couplings. If $N_1$ mass is fixed at $100$ GeV along with the Yukawa coupling $y$ such that the decay $\Gamma_{N_{1} \to \phi \chi}$ is of the order $\mathcal{O}(10^{-20})$, then one can obtain the required freeze-out cross-section of $N_1$ that can lead to the correct relic of SIDM by maintaining a mass-difference of $20-30$ GeV and $5-10$ GeV between $N_1$ and $\eta$ for $m_\chi=1$ GeV, $m_{\chi}=10$ GeV respectively.

\subsection{Gauged $B-L$ realization}
In this subsection, we consider another interesting UV completion of RHN portal SIDM scenario based on the gauged $B-L$ symmetry where $B$ and $L$ correspond to baryon and lepton numbers respectively. Since its proposal and early studies several decades ago \cite{Davidson:1978pm, Mohapatra:1980qe, Marshak:1979fm, Masiero:1982fi, Mohapatra:1982xz, Buchmuller:1991ce}, gauged $B-L$ scenarios have become a popular BSM framework as it addresses the problem of neutrino mass and, in some specific non-minimal realizations, can also provide a realistic dark matter candidate. Here we consider the minimal scenario based on gauged $B-L$ framework where, in addition to SM and SIDM sectors, three RHNs and one additional singlet scalar $\zeta$ are required. The BSM particle content is shown in table \ref{tab:tab3}. The three RHNs with $B-L$ charge $-1$ each also lead to cancellation of triangle anomalies keeping the model anomaly free.

\begin{table}[h!]
	\begin{tabular}{|c|c|c|c|}
		\hline \multicolumn{2}{|c}{Fields}&  \multicolumn{1}{|c|}{ $ SU(3)_c \otimes SU(2)_L \otimes U(1)_Y$ $\otimes~~U(1)_{\rm B-L} \otimes  U(1)_D $  } \\ \hline
		{Fermion} &  $N_R$&  ~~1 ~~~~~~~~~~~1~~~~~~~~~~0~~~~~~~~~~~-1~~~~~~~~~ 0 \\ 
		& $\chi$  & ~~1 ~~~~~~~~~~~1~~~~~~~~~~0~~~~~~~~~~~0~~~~~~~~~ 1 \\
		[0.5em] \cline{2-3}
		\hline
		Scalars & 
		$\Phi$ & ~~1 ~~~~~~~~~~~1~~~~~~~~~~0~~~~~~~~~~~1~~~~~~~~~ -1 \\
		& $\zeta$ & ~~1 ~~~~~~~~~~~1~~~~~~~~~~0~~~~~~~~~~~2~~~~~~~~~ 0 \\
		\hline
	\end{tabular}
\caption{BSM particle content and their transformation under the chosen symmetry in gauged $B-L$ realization of RHN portal SIDM.}
\label{tab:tab3}
\end{table}

The relevant Lagrangian for neutrino mass generation consistent with the imposed symmetry is given by
\begin{equation}
\mathcal{L} \supset -\frac{1}{2}f \zeta \overline{N^c_R} N_R - Y_\nu  \overline{L} \widetilde{H} N_R+ {\rm h.c.}
\end{equation}
The singlet scalar $\zeta$, after acquiring a non-zero VEV, denoted by $v_{BL}$, not only lead to spontaneous breaking of gauged $B-L$ symmetry but also generated RHN mass $M_{N_R}= f v_{BL}/\sqrt{2}$. And through the Yukawa coupling, after the EWSB, the neutrinos acquire a Dirac mass which is equal to $m_D = Y_\nu ~{v}/{\sqrt{2}}$. Since $N_R$ has Majorana mass $M_{N_R}$ and also mixes with light neutrinos via Dirac mass $m_D$, one needs to diagonalise the mass matrix in SM neutrino and RHN basis. Assuming a hierarchy $m_D \ll M_{N_R}$ leads to the light neutrino mass matrix via type-I seesaw mechanism as
\begin{equation}
M_\nu = -m_D M^{-1}_{N_R} m^T_D
\end{equation}
which is a $3 \times 3$ complex matrix and can be diagonalised by the PMNS mixing matrix in diagonal charged lepton basis. 

Now, as far as the lightest RHN $N_1$ acting as the portal to SIDM is concerned, there are two crucial difference between gauged $B-L$ model and scotogenic model discussed earlier. We elaborate them below.
\begin{itemize}
\item $N_1$ couples to SM leptons via SM Higgs $H$ and hence can not be made stable kinematically against decay into SM particles. This requires the tuning of $N_1$ Yukawa with SM leptons such that $N_1$ can decay dominantly into SIDM at late epochs. Such fine-tuning of $N_1$ Yukawa with SM leptons lead to vanishingly small lightest neutrino mass.
\item Since $N_1$ can have tiny couplings with the SM leptons, DM in $B-L$ model can, in principle, decay into SM neutrinos which can be constrained from different neutrino experiments, as discussed in \cite{Bandyopadhyay:2020qpn}. As a conservative approach, we consider $N_1$ coupling with leptons to be vanishingly small in our analysis.
\item Even though $N_1$ Yukawa with SM leptons is very small, $N_1$ can still be produced in the thermal bath if the gauged $B-L$ portal interactions are sizeable enough. Thus, even though the prospects of probing this scenario at LFV experiments are substantially less compared to the scotogenic realization, the gauged $B-L$ interactions can be probed at colliders.
\end{itemize}

\begin{figure}[h!]
\centering
\includegraphics[scale=0.5]{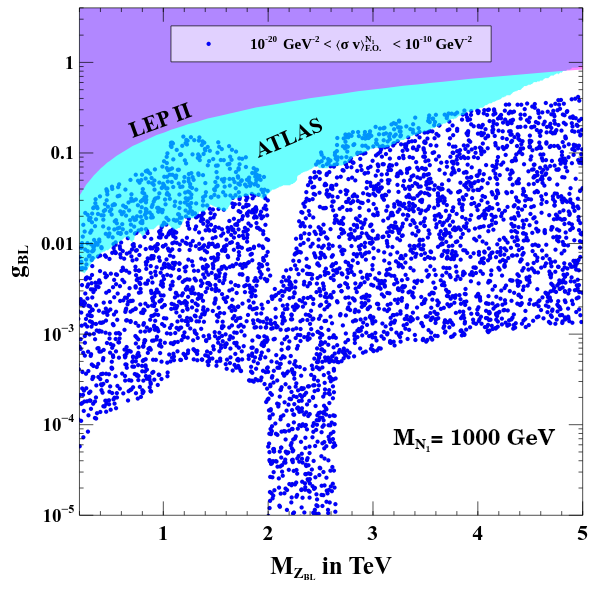}
\hfil
\includegraphics[scale=0.5]{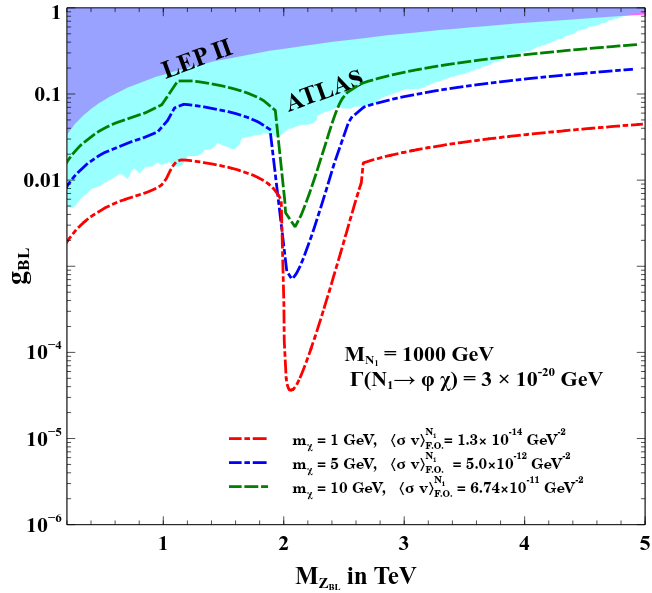}
\caption{Left: Parameter space for gauged $B-L$ model consistent with the required cross section of the lightest RHN in order to generate its required thermal relic, to be transferred to SIDM at late epochs.; Right: Contours of the required cross-section $\langle\sigma v \rangle^{^{N_1}}_{_{ \rm F.O.}}$ that can give rise to correct relic abundance for different DM masses ($m_{\chi} = 1,5,10$ GeV) with a fixed $N_1$ mass ($M_{N_1}=1000$ GeV) and decay width ($\Gamma_{N_{1} \to \phi \chi}=3\times10^{-20}$ GeV) in the plane of $M_{Z_{\rm BL}}$ and $g_{\rm BL}$. }
\label{blps}
\end{figure}

In general both gauged $B-L$ and scalar portal interactions can be responsible for keeping the RHNs in thermal equilibrium with the SM bath in early epochs. The scalar portal interactions can arise due to mixing of $\zeta$ with the SM Higgs which we ignore in our work and constrain the gauge interactions only from the requirement of appropriate cross-section $\langle \sigma v \rangle_{N_1 N_1 \rightarrow {\rm SM\, SM}}$ leading to a freeze-out abundance of $N_1$. The gauge interactions of $N_1$ arise due to the kinetic terms as
\begin{equation}
\mathcal{L}_{\rm kinetic} \supset i\overline{{N}_{R}}
\slashed{D}{N}_{R}, \,\,\,\, D_{\mu} = \partial_{\mu}-ig_{\rm BL} (Z_{\rm BL})_{\mu}.
\end{equation}
At late epochs, $N_1$ can decay into SIDM $\chi$ to fulfill the relic criteria of the latter as discussed in subsection~\ref{dm_production}.
In left panel of Fig.~\ref{blps}, we show the parameter space in the plane of $\rm B-L$ gauge coupling $g_{\rm BL}$ and the gauge boson mass $M_{Z_{\rm BL}}$ consistent with the required $N_1$ annihilation cross-section or freeze-out abundance. For the scan $N_1$ mass was fixed at $1000$ GeV and $M_{Z_{\rm B-L}}$ was varied upto $5$ TeV. Clearly we can see the resonance feature around $M_{Z_{\rm BL}}=2$ TeV due to dominant $N_1$ annihilation mediated by $Z_{\rm BL}$. In the right panel of Fig.~\ref{blps}, we show the contours of the required $N_1$ annihilation cross-section in the plane of $M_{Z_{\rm BL}}$ and $g_{\rm BL}$ that can give rise to correct relic abundance for different DM masses ($m_{\chi} = 1,5,10$ GeV) with a fixed $N_1$ mass ($M_{N_1}=1000$ GeV) and decay width ($\Gamma_{N_{1} \to \phi \chi}=3\times10^{-20}$ GeV). 


While we need sizeable $B-L$ gauge sector coupling with light $Z_{\rm BL}$ so that $N_1$ can attain the required freeze-out relic, the corresponding parameter space is tightly constrained from collider experiments. The limits from LEP II data constrains such additional gauge sector by imposing a lower bound on the ratio of new gauge boson mass to the new gauge coupling $M_{Z'}/g' \geq 7$ TeV \cite{Carena:2004xs, Cacciapaglia:2006pk}. The bounds from ongoing LHC experiment have already surpassed the LEP II bounds. In particular, search for high mass dilepton resonances have put strict bounds on such additional gauge sector coupling to all generations of leptons and quarks with coupling similar to electroweak ones. The latest bounds from the ATLAS experiment \cite{Aaboud:2017buh, Aad:2019fac} and the CMS experiment \cite{Sirunyan:2018exx} at the LHC rule out such gauge boson masses below 4-5 TeV from analysis of 13 TeV data. Such bounds get weaker, if the corresponding gauge couplings are weaker \cite{Aaboud:2017buh} than the electroweak gauge couplings. We show these exclusion limits in Fig.~\ref{blps}. Clearly, the ATLAS limit from 13 TeV LHC data rules out some portion of the parameter space while keeping the other parts predictive at near future runs.

Here it is worth mentioning that, for a particular DM mass $m_\chi$, if $\Gamma_{N_{1} \to \phi \chi}$ is further decreased (Mass of $N_1$, $M_{N_{1}}$ and Yukawa coupling $y$ can always be tuned accordingly) upto $10^{-21}$ GeV, then that would require a larger freeze-out cross-section $\langle\sigma v \rangle^{^{N_1}}_{_{ \rm F.O.}}$ so as to give rise to correct relic abundance of DM $\chi$ which can be infered from Fig.~\ref{relic_scan}. To obtain a larger freeze-out cross-section through the ${\rm B-L}$ gauge portal, one would require to have a larger gauge coupling $g_{\rm BL}$. But since such larger gauge couplings are constrained from the collider experiments, so in that case some more parameter space will get ruled out. Beyond $\Gamma_{N_{1} \to \phi \chi}=10^{-21}$ GeV, if it is further decreased, then there would be no effect on the $g_{\rm BL}-M_{Z_{\rm BL}}$ parameter space, as the same freeze-out cross-section can lead to correct relic of $\chi$ which is explained clearly in section~\ref{dm_production}. On the contrary, if $\Gamma_{N_{1} \to \phi \chi}$ is increased, then the required $\langle\sigma v \rangle^{^{N_1}}_{_{ \rm F.O.}}$ will decrease which can be easily achieved through the $\rm B-L$ gauge portal without any tension from the collider constraints from ATLAS and LEP II. 
\section{Conclusion}
\label{sec:conclude}
We have studied a self-interacting dark matter scenario consistent with astrophysical requirements of addressing the small scale issues of cold dark matter paradigm. Due to the existence of a light mediator, considered to be the vector boson of a gauged dark Abelian symmetry, velocity-dependent DM self-interactions can be naturally realised giving rise to the required differences across astrophysical scales from dwarf galaxies to clusters. Depending upon the kinetic mixing of dark $U(1)_D$ with the SM hypercharge $U(1)_Y$, DM can be produced either thermally or non-thermally from the SM bath although the thermal scenario faces tight constraints from DM direct detection experiments. Irrespective of thermal or non-thermal production, the final DM relic remains sub-dominant due to strong DM annihilations into its light mediators by virtue of large self-interaction coupling. In order to fill the relic deficit of SIDM, we propose a right handed neutrino portal SIDM scenario which can also have non-trivial connection to the origin of light neutrino masses depending upon the particular UV completion.

We first show the salient features of this scenario by discussing the minimal model along with DM self-interactions, DM relic via a hybrid of thermal and non-thermal mechanisms and DM direct detection. We then propose two UV complete realizations of this minimal RHN portal SIDM setup namely, scotogenic and gauged $B-L$ realizations. In scotogenic realization, in addition to SM and SIDM sectors, there exist three RHNs and an extra scalar doublet odd under an unbroken $Z_2$ symmetry. While light neutrino mass arises at one-loop with $Z_2$-odd particles going inside the loop, the lightest RHN $N_1$ can acquire a freeze-out relic due to sizeable interactions with SM leptons. While $N_1$ decay into SM lepton is forbidden kinematically, it can decay into SIDM at late epochs thereby generating the required relic. The requirement of sizeable Yukawa couplings of $N_1$ also keeps the model predictive at experiments looking for charged lepton flavour violation. As a second example of UV completion, we consider another popular framework based on the gauged $B-L$ symmetry where three RHNs are also required as a minimal solution to the triangle anomaly cancellation, apart from their roles in generating light neutrino masses via type-I seesaw mechanism. While the lightest RHN decay into SM leptons can be made negligible by appropriate fine-tuning of Dirac Yukawa couplings with the consequence of vanishingly small lightest neutrino mass, the required $N_1$ freeze-out relic (to fill the SIDM relic deficit via late decay) can still be generated by virtue of gauged $B-L$ interactions. While the LFV prospects are low due to fine-tuned Yukawa of the lightest RHN with leptons, the model remains predictive at collider experiments due to $B-L$ interactions. Additionally, the $B-L$ framework also allows for the option of long-lived DM as there is no unbroken symmetry protecting its stability. While we did not pursue the indirect detection phenomenology of such a scenario and kept the $N_1$ coupling to leptons vanishingly small, such long-lived DM decaying into neutrinos can be probed at different neutrino detectors. Additionally, keeping $N_1$-lepton coupling non-vanishing but comparable to $N_1$-DM coupling will also involve a more rigorous treatment of relic density calculation as $N_1$ can simultaneously convert into radiation and DM at late epochs. We leave such interesting possibilities to future works. Finally, to summarize our present work, the minimal setup of RHN portal SIDM remains predictive with respect to observations in astrophysics, cosmology and DM direct detection experiments while the specific UV completions can offer complementary probes at experiments in energy and intensity frontiers.

\acknowledgements
DB acknowledges the support from Early Career Research Award from Science and Engineering Research Board (SERB), Department of Science and Technology (DST), Government of India (reference number: ECR/2017/001873). MD acknowledges DST, Government of India for providing the financial assistance for the research under the grant 
DST/INSPIRE/03/ 2017/000032.
\appendix	
\section{Relevant cross section and decay widths}	
\label{appen1}
\begin{equation}
\begin{aligned}
	\Gamma({N_R\rightarrow \phi \chi}) &= \frac{y^2}{32\pi M^3_{N_R}}\big((M_{N_R} + m_\chi)^2 - m^2_\phi \big)\\
	&\big( M^4_{N_R} + m^4_\chi + m^4_\phi - 2 M^2_{N_R} m^2_{\chi} - 2 m^2_\chi m^2_\phi - 2 M^2_{N_R} m^2_\phi \big)^{\frac{1}{2}}
\end{aligned}
\end{equation}
\begin{equation}
\begin{aligned}		
	\sigma({\rm \chi \;\chi} \rightarrow Z' Z') &= \frac{g'^4}{192 \pi s (s-4m^2_{\chi})} \times \Bigg[\frac{24s(4m^4_{\chi}+2M^4_{Z'}+sm^2_{\chi})A}{M^4_{Z'}+m^2_{\chi}s-4M^2_{Z'}m^2_{\chi}}\\ & -\frac{24 (8m^2_{\chi}-4M^2_{Z'}-s^2-(s-2M^2_{Z'})4m^2_{\chi})}{s-2M^2_{Z'}} {\rm Log}\Big[\frac{2M^2_{Z'}+s(A-1)}{2M^2_{Z'}-s(A+1)}\Big]\Bigg]
\end{aligned}
\end{equation}
where $A=\sqrt{\frac{(s-4M^2_{Z'})(s-4m^2_{\chi})}{s^2}}$
\begin{eqnarray}
\sigma( e^{+} e^{-} \rightarrow {\rm \chi \;\chi})&=& \frac{g^2 g'^2 \epsilon^2(s+2m^2_{\chi})(s-m^2_e-4(s+2m^2_e)\sin^2\theta_{W})}{96 \pi \cos^2\theta_{W}(s-4m^2_e)(s-m^2_{Z'})^2}\sqrt{\frac{(s-4m^2_e)(s-4m^2_{\chi})}{s^2}} \nonumber \\
\end{eqnarray}	

Thermal averaged cross-section for annihilation of any particle $A$ to $B$  is given by: \cite{Gondolo:1990dk}
\begin{equation}
\langle\sigma v \rangle_{AA \rightarrow BB} = \frac{x}{2\big[K^2_1(x)+K^2_2(x)\big]}\times \int^{\infty}_{2}  dz \sigma_{(AA\rightarrow BB)} (z^2 m^2_A) (z^2 - 4)z^2 K_1(zx)
\label{appeneq1}
\end{equation}
where $z=\sqrt{s}/m_A$ and $x=m_A/T$.

Thermal averaged decay width of $\Phi_1$ decaying to $\chi_1$ is given by:
\begin{equation}
\langle \Gamma(\Phi_1 \rightarrow \chi_1 \chi_1) \rangle = \Gamma(\Phi_1 \rightarrow \chi_1 \chi_1)  \Bigg(\frac{K_1(x)}{K_2(x)}\Bigg)
\label{appeneq2}
\end{equation}

In Eqn. \eqref{appeneq1} and \eqref{appeneq2}, $K_1$ and $K_2$ are the modified Bessel functions of 1st and 2nd kind respectively.

The decay with of $Z'$ into a $e^+ e^{-}$ pair or pair of neutrinos:
\begin{equation}
\Gamma_{Z\to f \bar{f}} = \frac{\epsilon^2 g^2 M_{Z'}}{48 \pi \cos^2\theta_{w}}\big(C^2_{f_A} + C^2_{f_V}\big)
\end{equation}

\section{Loop Functions}
\label{loopfunc}
The loop functions used in section~\ref{lfv} are given by \cite{Toma:2013zsa}
\begin{eqnarray}
	F_2(x) &=& \frac{1-6x+3x^2+2x^3-6x^2 \log x}{6(1-x)^4}, \\
	G_2(x) &=& \frac{2-9x+18x^2-11x^3+6x^3 \log x}{6(1-x)^4}, \\
	D_1(x,y) &=& - \frac{1}{(1-x)(1-y)} - \frac{x^2 \log x}{(1-x)^2(x-y)} -
	\frac{y^2 \log y}{(1-y)^2(y-x)}, \\
	D_2(x,y) &=& - \frac{1}{(1-x)(1-y)} - \frac{x \log x}{(1-x)^2(x-y)} -
	\frac{y \log y}{(1-y)^2(y-x)}.
\end{eqnarray}
These loop functions do not have any poles. In the limit $x,y\to1$ and
$y\to x$, the functions become 
\begin{eqnarray}
	F_2(1)&=&\frac{1}{12},\quad
	G_2(1)=\frac{1}{4},\quad
	D_1(1,1)=-\frac{1}{3},\quad
	D_2(1,1)=\frac{1}{6},
\end{eqnarray}
\begin{eqnarray}
	&&D_1(x,x)=\frac{-1+x^2-2x\log{x}}{(1-x)^3},\\
	&&D_1(x,1)=D_1(1,x)=\frac{-1+4x-3x^2+2x^2 \log{x}}{2(1-x)^3},\\
	&&D_2(x,x)=\frac{-2+2x-(1+x)\log{x}}{(1-x)^3},\\
	&&D_2(x,1)=D_2(1,x)=\frac{1-x^2+2x\log{x}}{2(1-x)^3}.
\end{eqnarray}

\section{Couplings relevant for $\pmb{\mu \to e}$ conversion in nuclei}
\label{appen3}

In the above, $g_{XK}^{(0)}$ and $g_{XK}^{(1)}$ (with $X = L, R$ and $K = S, V$) are given by
\begin{align}
	g_{XK}^{(0)} &= \frac{1}{2} \sum_{q = u,d,s} \left( g_{XK(q)} G_K^{(q,p)} +
	g_{XK(q)} G_K^{(q,n)} \right)\,, \nonumber \\
	g_{XK}^{(1)} &= \frac{1}{2} \sum_{q = u,d,s} \left( g_{XK(q)} G_K^{(q,p)} - 
	g_{XK(q)} G_K^{(q,n)} \right)\,.
\end{align}
Neglecting the Higgs-penguin contributions due to the smallness of the
involved Yukawa couplings. Therefore, the corresponding couplings are
\begin{eqnarray}
	g_{LV(q)} &=& g_{LV(q)}^{\gamma} + g_{LV(q)}^{Z}\,, \nonumber \\
	g_{RV(q)} &=& \left. g_{LV(q)} \right|_{L \leftrightarrow R}\,, \nonumber \\
	g_{LS(q)} &\approx& 0 \, , \nonumber \\ 
	g_{RS(q)} &\approx& 0 \, .
\end{eqnarray}
The photon and $Z$-boson couplings can be computed from the Feynman
diagrams which are given by:
\begin{align}
	g_{LV(q)}^{\gamma} &= \frac{\sqrt{2}}{G_F} e^2 Q_q 
	\left(A_{ND} - A_D \right)\,, \nonumber \\
	g_{RV(q)}^{Z} &= -\frac{\sqrt{2}}{G_F} \, \frac{g_L^q + g_R^q}{2} \, 
	\frac{F}{M_Z^2} \,. 
\end{align}
And the tree-level $Z$-boson couplings to a pair of quarks are:
\begin{equation}
	g_L^q = \frac{g_2}{\cos \theta_W}\left( Q_q \sin^2 \theta_W - T_3^q
	\right), \qquad g_R^q = \frac{g_2}{\cos \theta_W} Q_q \sin^2 \theta_W,
\end{equation}

\bibliographystyle{JHEP}
\bibstyle{JHEP}
\providecommand{\href}[2]{#2}\begingroup\raggedright\endgroup

\end{document}